\documentclass{jfm}
\usepackage{amsmath}
\usepackage{amsfonts}
\usepackage{amssymb}
\usepackage{amsmath}
\usepackage{graphicx}
\usepackage{natbib}
\usepackage{float}
\usepackage{subfig}
\usepackage{comment}
\usepackage{ulem}

\usepackage{lineno}
\usepackage{epstopdf}
\usepackage{soul}
\usepackage{color}

\title[Interactions between near-inertial waves and mean flow]{A generalised-Lagrangian-mean model of the interactions between near-inertial waves and mean flow}
\author{J.-H. Xie and J. Vanneste}
\affiliation{School of Mathematics and Maxwell Institute for Mathematical Sciences, \\ University of Edinburgh, Edinburgh EH9 3FD, UK}
\date{\today}

\def\bx{\boldsymbol{x}}
\def\bX{\boldsymbol{X}}
\def\ba{\boldsymbol{a}}
\def\bu{\boldsymbol{u}}
\def\d{\mathrm{d}}
\def\bxi{\boldsymbol{\xi}}
\def\cc{\mathrm{c.c.}}
\def\eps{\epsilon}

\newcommand{\av}[1]{\langle #1 \rangle}
\newcommand{\mr}[1]{\mathrm{#1}}
\newcommand{\br}[1]{\left( #1 \right)}
\newcommand{\ex}{\mathrm{e}}
\newcommand{\ii}{\mathrm{i}}
\newcommand{\Ro}{\textrm{Ro}}
\newcommand{\jaco}[6]{\frac{\partial(#1,~#2,~#3)}{\partial(#4,~#5,~#6)}}
\newcommand{\mc}[1]{\mathcal{#1}}
\newcommand{\abs}[1]{\left| #1 \right|}
\newcommand{\bs}[1]{\boldsymbol{#1}}
\newcommand{\jacos}[4]{\frac{\partial(#1,~#2)}{\partial(#3,#4)}}
\newcommand{\eqn}[1]{(\ref{#1})}

\newcommand*\patchAmsMathEnvironmentForLineno[1]{%
  \expandafter\let\csname old#1\expandafter\endcsname\csname #1\endcsname
  \expandafter\let\csname oldend#1\expandafter\endcsname\csname end#1\endcsname
  \renewenvironment{#1}%
     {\linenomath\csname old#1\endcsname}%
     {\csname oldend#1\endcsname\endlinenomath}}%
\newcommand*\patchBothAmsMathEnvironmentsForLineno[1]{%
  \patchAmsMathEnvironmentForLineno{#1}%
  \patchAmsMathEnvironmentForLineno{#1*}}%
\AtBeginDocument{%
\patchBothAmsMathEnvironmentsForLineno{equation}%
\patchBothAmsMathEnvironmentsForLineno{align}%
\patchBothAmsMathEnvironmentsForLineno{flalign}%
\patchBothAmsMathEnvironmentsForLineno{alignat}%
\patchBothAmsMathEnvironmentsForLineno{gather}%
\patchBothAmsMathEnvironmentsForLineno{multline}%
}

\begin{document}
\maketitle

\begin{abstract}
Wind forcing of the ocean generates a spectrum of inertia-gravity waves that is sharply peaked near the local inertial (or Coriolis) frequency. The corresponding near-inertial waves (NIWs) are highly energetic and play a significant role in the slow, large-scale dynamics of the ocean. To analyse this role, we develop a new model of the non-dissipative interactions between NIWs and balanced motion. The model is derived using the generalised-Lagrangian-mean (GLM) framework (specifically, the glm variant of \citet{Sowa2010}), taking  advantage of the time-scale separation between the two types of motion to average over the short NIW period. 

We combine \citeauthor{Salm2013}'s (\citeyear{Salm2013}) variational formulation of GLM with Whitham averaging to obtain a system of equations governing the joint evolution of  NIWs  and mean flow. Assuming that the mean flow is  geostrophically balanced reduces this system to a simple model coupling \citeauthor{Youn1997}'s (\citeyear{Youn1997}) equation for NIWs  with a modified quasi-geostrophic equation. In this coupled model, the mean flow affects the NIWs through advection and refraction; conversely, the NIWs affect the mean flow by modifying the potential-vorticity inversion --  the relation between advected potential vorticity and advecting mean velocity -- through  a quadratic wave term, {consistent with the GLM results of \citet{Buhl1998}.}

The coupled model is Hamiltonian and its conservation laws, for wave action and energy in particular, prove illuminating: on their basis, we identify a new interaction mechanism whereby NIWs forced at large scales extract energy from the balanced flow as their horizontal scale is reduced by differential advection and refraction so that their potential energy increases. A rough estimate suggests that this mechanism could provide a significant sink of energy for mesoscale motion and play a part in the global energetics of the ocean.

Idealised two-dimensional models are derived and simulated numerically to gain insight into NIW--mean-flow interaction processes. A simulation of a one-dimensional barotropic jet demonstrates how NIWs forced by wind slow down the jet as they propagate into the ocean interior. A simulation assuming plane travelling NIWs in the vertical shows how a vortex dipole is deflected by NIWs, illustrating the irreversible nature of the interactions. In both simulations energy is transferred from the mean flow to the NIWs. 
 
\end{abstract}

\section{Introduction}

Near-inertial waves (NIWs), that is, inertia-gravity waves with frequencies close to the local Coriolis frequency $f_0$, play an important role in the dynamics of the ocean \citep[e.g.][]{Fu1981}. They account for almost 50\% of the wave energy \citep[e.g.][]{Ferr2009} and thus make a strong contribution to processes associated with inertia-gravity waves such as diapycnal mixing, vertical motion and  primary production. Several features explain their dominance \citep{Garr2001}: their minimum frequency in the inertia-gravity-wave spectrum, the low frequency of the atmospheric winds that generate them, the presence of turning latitudes, nonlinear interactions \citep{Medv2007}, and the transfer of tidal energy through parametric subharmonic instability
\citep{Youn2008}.

In view of their large energy, it is natural to expect that NIWs affect the large-scale circulation of the ocean. One possibility is that they do so through enhanced diapycnal mixing in the regions of the ocean where they dissipate \citep[e.g.][]{Wuns2004}. Another, more remarkable perhaps, is that they  alter the slow, balanced oceanic circulation directly through wave--mean-flow interaction processes. \citet{Gert2009} put forward the idea that NIWs  provide an energy sink for this circulation. Their numerical simulations suggest that this process may be significant and, along with other mechanisms including bottom and surface friction \citep[e.g.][]{Niku2013,Duha2006} and loss of balance \citep[e.g.][]{Vann2013,dani-et-al}, help resolve the long-standing puzzle posed by the  dissipation of the (inverse energy-cascading) balanced oceanic flow. 

The aim of the present paper is to develop a theoretical tool that enables a detailed analysis of the interactions between NIWs and balanced flow. So far, theoretical modelling has focussed on the impact of the balanced flow on NIWs. 
Under the assumption that NIW scales are much smaller than mean-flow scales, a WKB approach can be applied \citep{Mooers1975a,Mooers1975b}; it shows in particular that the vorticity of the balanced flow shifts the frequency of NIWs away from $f_0$  \citep{Kunz1985}. \citet{Youn1997} (referred to as YBJ hereafter) derived an asymptotic model based on the frequency separation between NIWs and balanced motion which, in contrast, makes no assumption of {separation between the NIW and flow spatial scales}. Their model is therefore well suited to examine the realistic scenario of NIWs forced by atmospheric winds at horizontal scales larger than those of the ocean flow. The YBJ model describes the slow modulation of the NIW fields about their oscillations at the fast frequency $f_0$. It neatly isolates the main mechanisms whereby the balanced flow and stratification influence NIWs: advection, dispersion and refraction. 

In this paper, we extend the YBJ model to account for the feedback of the NIWs on the balanced flow. Specifically, we derive a new model that couples the YBJ model with a modified quasi-geostrophic (QG) model. The modification -- a change in the relation between the advected potential vorticity (PV) and advecting velocity {involving quadratic wave terms} -- captures this feedback. 
As detailed below, we work in the framework of non-dissipative generalised Lagrangian-mean theory \citep[GLM, see, e.g.,][]{buhl09}. {\citeauthor{Buhl1998} (1998; for short waves) and  \citeauthor{Holm2011} (2011; for waves of arbitrary spatial scales) showed that the change in the PV--velocity relationship is a generic conclusion of this theory which interprets some of the quadratic wave terms as a PV contributions associated with the wave pseudomomentum  \citep[see also][]{Buhl2005,buhl09,Salm2013}.}

{We pay close attention to the conservation laws satisfied by the coupled model.
These} turn out to be particularly important: based on the conservation of NIW action (in fact, the NIW kinetic energy divided by $f_0$)  and total energy alone, we identify a novel mechanism providing a sink of energy for the balanced flow. In this mechanism, the reduction in the horizontal scale of  NIWs that results from  advection and refraction is accompanied by an increase in the NIW potential energy and, consequently, a decrease in the energy of the balanced flow.

%

A key to the derivation of wave--mean-flow  models of the kind we develop is to separate the motion between mean and wave contributions, relying on the  time-scale separation to define the mean as an average over the inertial period $2\pi/f_0$. The GLM theory of \citet{Andr1978} offers a general framework for this separation and for the systematic derivation of equations governing the coupled wave--mean dynamics (see \citealt{buhl09} for an account). The theory has achieved notable successes but suffers from a deficiency in that the (Lagrangian) mean velocity it defines is divergent even for an incompressible fluid. \citet{Sowa2010} proposed a variant of GLM, termed `glm', which yields a divergence-free mean velocity. Because it is convenient, we adopt this approach in the main body of the paper but show in an  Appendix that the same leading-order model can also be obtained from standard GLM. We also adopt a variational approach that ensures that conservation laws and their link to symmetries are preserved when the primitive equations are reduced asymptotically (see, e.g., \citealt{Salm1988,Grim1984,Holm2009}). Specifically, we derive the Lagrangian-mean and perturbation equations by introducing a wave--mean decomposition of the flow map into the primitive-equation Lagrangian, following closely the method proposed by  \citet{Salm2013} (see \citealt{Gjaj1996} for a related approach). 
Because the wave component consists of rapidly oscillating NIWs, the resulting Lagrangian can be averaged in time in the manner of \citet{Whit1974}. Variations with respect to the mean flow map (or rather its inverse) and to the NIW amplitude then lead to a coupled primitive-equation--YBJ system; applying a QG approximation  reduces this system to a simple, energy conserving YBJ-QG coupled model. (See \citet{Vann2014} for a related variational derivation of the original YBJ equation.)

The paper is organised as follows. The coupled YBJ-QG model is introduced without a derivation in \S\,\ref{sec:coupledmodel}. Some key properties of the model and the key scaling assumptions underlying its derivation are also discussed there. The derivation itself is carried out in \S\,\ref{sec:derivation} which also records the complete primitive-equation--YBJ model. The Hamiltonian structure of the YBJ-QG model and associated conservation laws are presented in \S\,\ref{sec:hamilton}. Sections \ref{sec:derivation} and \ref{sec:hamilton} are technical; the reader mainly interested in applications can skip them and move directly to \S\,\ref{sec:implications} which considers the 
possible implications of the wave--mean-flow interactions represented in the model for ocean energetics. Section \ref{sec:2d} examines two simplified models deduced from the full YBJ-QG model assuming certain symmetries. These models are two-dimensional and hence easily amenable to numerical simulations. We take advantage of this and present the results of two sets of simulations demonstrating (i) the slow down of a one-dimensional barotropic jet by NIWS, and (ii) the deflection of a vortex dipole under the influence of vertically travelling NIWs. The paper concludes with a brief Discussion in \S\,\ref{sec:discussion}. 
Three Appendices provide details of some of the computations and alternative derivations.

\section{Coupled model} \label{sec:coupledmodel}

\subsection{Model}

We start with the hydrostatic--Boussinesq equations written in the form
\begin{subequations} \label{hydrobou}
\begin{align}
\partial_t u + \bu \cdot \nabla u + w \partial_z u - (f_0+\beta y) v &= - \partial_x p, \label{Dyn_Equ_1}\\
\partial_t v + \bu \cdot\nabla v + w\partial_z v + (f_0+\beta y) u &= - \partial_y p, \label{Dyn_Equ_1.5}\\
\theta &= \partial_z p, \label{Dyn_Equ_2} \\
\nabla \cdot  \bu + \partial_z w &= 0, \label{incomp} \\
\partial_t \theta + \bu\cdot \nabla \theta + w\partial_z\theta  &= 0, \label{Dyn_Equ_n}
\end{align}
\end{subequations}
where $\bu=(u,v)$ is the horizontal velocity, $w$ is the vertical velocity, $p$ is the pressure, and $\theta$ is the buoyancy, defined as $-g$ times the density variations relative to a constant density $\rho_0$ \citep[e.g.][]{vall06}. We have used the $\beta$-plane approximation to write the Coriolis parameter as $f_0 + \beta y$, with constant $f_0$ and $\beta$. Throughout the paper, $\nabla = (\partial_x,\partial_y)$ denotes the horizontal gradient.

Inertial oscillations are characterised by a linear balance between inertia and the Coriolis force in \eqn{Dyn_Equ_1}--\eqn{Dyn_Equ_1.5} and thus satisfy
\begin{equation}
\partial_t u - f_0 v = 0 \quad \textrm{and} \quad \partial_t v + f_0 u = 0. 
\end{equation}
The solution can written in complex form as
\begin{equation} \label{uiv}
u + \ii v =  M_z \, \ex^{-\ii f_0 t}
\end{equation}
for some complex amplitude $M(x,y,z)$. Here we follow YBJ in writing this amplitude as a $z$-derivative so that the vertical velocity, deduced from the incompressibility condition \eqn{incomp}, takes the simple form
\begin{equation} \label{w}
w = - M_s \ex^{-\ii f_0 t} + \cc,
\end{equation}
where $s=x+\ii y$, $\partial_{s} = (\partial_x-\ii \partial_y)/2$, and $\cc$ denotes the complex conjugate of the preceding term. 
The position $\bx=(x,y,z)$ of fluid particles in the inertial field \eqn{uiv}--\eqn{w} can be
obtained by integration. If this position is written as
\begin{equation}
\bx = \bX + \bxi,
\end{equation}
the displacement $\bxi=(\xi,\eta,\zeta)$ satisfies 
\begin{equation}
\xi + \ii \eta = \chi^{}_z \, \ex^{-\ii f_0 t} \quad \textrm{and} \quad \zeta = - \chi^{}_s \ex^{-\ii f_0 t} + \cc \label{displace1}
\end{equation}
where $\chi^{} = \ii M/f_0$ in the linear approximation. The mean position $\bX$ can be regarded as an integration constant identifying the fluid particle, and the displacement $\bxi$ and amplitude $\chi$ can be thought of as functions of $\bX$.

For NIWs propagating in a flow, the description leading to \eqn{displace1} is overly simplified. However, it can be extended to capture the two-way interactions between the NIWs and the flow: this is achieved by regarding $\bX$ as a suitably defined, time-dependent Lagrangian-mean position (in fact, a mean map $\bX(\ba,t)$ mapping the particle labelled by $\ba$ to its mean position at time $t$), and by taking the amplitude $\chi(\bX,t)$ to be a function of both time and mean position in typical GLM fashion \citep[e.g.][]{buhl09}. The main achievement of this paper is the derivation of equations governing the joint evolution of the NIW amplitude $\chi$ and of the mean map $\bX(\ba,t)$ or, rather, of the corresponding Lagrangian-mean velocity. 

We leave the details of this derivation for the next section and present here the final equations. These are particularly simple when the Lagrangian-mean flow is assumed to be quasi-geostrophic and hence derived from a streamfunction $\psi$ according to $(\bar{\bu}^\mathrm{L},\bar w^\mathrm{L})=(\nabla^\bot \psi,0)$, with $\nabla^\bot=(-\partial_y,\partial_x)$. In this approximation, and using $\bx$ rather than $\bX$ to denote the independent spatial variables (the mean positions),  the coupled model takes the form 
\begin{subequations}\label{model}
\begin{align}
& \chi^{}_{zzt} + \left(\partial(\psi,\chi^{}_z)\right)_z + \ii \beta y \chi^{}_{zz} \nonumber \\
& \qquad \qquad  + \frac{\ii}{2} \left( \left(\frac{N^2}{f_0} + \psi_{zz} \right) \nabla^2 \chi^{} + \nabla^2 \psi \,\chi^{}_{zz}  - 2\nabla \psi_z\cdot \nabla \chi^{}_z \right) = 0, \label{YBJ} \\
& q_t + \partial(\psi,q) = 0, \label{MQGPV_Eq}
\end{align}
\end{subequations}
where $\partial(\cdot,\cdot)$ denotes the two-dimensional Jacobian (with $\partial(f,g)=f_x g_y -  g_x f_y$),  and 
$N$ is the Brunt--V\"ais\"al\"a frequency which generally depends on $z$ and { is defined by $N^2=\bar \theta_z$ with $\bar \theta$ the background stratification.}

The first equation can be recognised as a version of the YBJ model, specifically their complete Eq.\ (3.2) rather than the simplified model given by their Eq.\ (1.2). It is supplemented by the boundary conditions at the top and bottom boundaries $z=z^\pm$,
\begin{equation}
\chi = \textrm{const}^\pm \  \ \textrm{at} \ \  z= z^\pm,
\end{equation}
ensuring a vanishing NIW vertical velocity there.
The second equation is the material conservation of the quasi-geostrophic potential vorticity (QGPV) $q$. This is related to the streamfunction $\psi$ and to $\chi$ through
\begin{equation}
q = \beta y + \Delta \psi + \frac{\ii f_0}{2} \partial(\chi_z^*,\chi^{}_z) + f_0 G(\chi^*,\chi^{}), \label{mqgpv}
\end{equation}
where 
\begin{equation}
\Delta = \nabla^2 + \partial_z \left(f_0^2/N^2 \partial_z\right), \label{Delta}
\end{equation}
is the familiar quasi-geostrophic potential vorticity operator,  
\begin{equation}
G(\chi^*,\chi) = \frac{1}{4}\br{ 2|\nabla \chi^{}_z|^2 - \chi^{}_{zz} \nabla^2\chi^* - \chi_{zz}^* \nabla^2\chi }, \label{G}
\end{equation}
and $^*$ denotes complex conjugate. In a familiar way, \eqn{mqgpv} should be interpreted as an inversion equation which relates the streamfunction $\psi$ and hence the advecting velocity $\nabla^\bot \psi$ to the dynamical variables, here $q$ and $\chi$. This inversion necessitates boundary conditions. In the vertical they are provided by the advection of the Lagrangian-mean buoyancy at the top and bottom boundaries, that is,
\begin{equation}
\partial_t \theta^\pm + \partial(\psi^\pm, \theta^\pm) =0, \quad \textrm{where} \ \ \psi^{\pm} = \left. \psi\right|_{z=z^\pm} \ \ \textrm{and} \ \ \theta^\pm = f_0 \left. \psi_z \right|_{z=z^\pm}. \label{Boun_Dyn}
\end{equation}

For horizontally periodic or unbounded domains, as assumed in what follows, Eqs.\ \eqn{model}--\eqn{Boun_Dyn} define the new model completely. The YBJ equation \eqn{YBJ} describes the weak dispersion that arises from a finite horizontal scale (through the term $\ii N^2 \nabla^2 \chi/(2f_0))$ and as well the various effects that the mean flow has on the NIWs: advection (term $(\partial(\psi,\chi^{}_z))_z$), and refraction by the mean vorticity (term $\ii \nabla^2 \psi  \, \chi^{}_{zz}/2$) and by vertical shear (term $- \ii \nabla \psi_z \cdot \nabla \chi^{}_z$). The simple QGPV equation \eqn{MQGPV_Eq} governs the mean flow. Here the effect of the NIWs is a modification of the relation between $\psi$ and $q$ by the quadratic wave terms in \eqn{mqgpv}. 
This structure is expected from GLM theory which interprets the quadratic wave terms as a potential vorticity contribution stemming from the wave pseudomomentum {\citep{Buhl1998,Holm2011}}. 

\subsection{Some properties}

An important feature of the coupled model is its conservation laws. The model conserves the total energy
\begin{equation}
\mc{H} = \frac{1}{2} \int \left( |\nabla\psi|^2 + \frac{f_0^2}{N^2}\psi_z^2 + f_0\beta y |\chi^{}_z|^2 + \frac{N^2}{2}\abs{\nabla \chi}^2 \right) \, \d\bx, \label{TotEne}
\end{equation}
and the wave action
\begin{equation}
\mc{A} = \frac{f_0}{2} \int |\chi^{}_z|^2 \, \d\bx. \label{Cons_Wave_Act}
\end{equation}
The wave action can be recognised as the kinetic energy of the NIWs divided by $f_0$. Its conservation does not follow from an analogous conservation in the hydrostatic--Boussinesq equations; rather it stems from an adiabatic invariance associated with the large time-scale separation between the fast oscillations of the NIWs and the slow evolution of their amplitude and of the mean flow \citep[cf.][]{Cotter-Reich}. Since, in the NIW limit, the leading-order wave energy is entirely kinetic and their frequency is $f_0$, the familiar form of wave action, namely the ratio of  wave energy to frequency, reduces to \eqn{Cons_Wave_Act}. The conservation of $\mc{H}$ is directly inherited from the energy conservation for the hydrostatic--Boussinesq equations. The first two terms in \eqn{TotEne} are recognised as the quasi-geostrophic kinetic and potential energy associated with the mean flow. The third term is associated with the $\beta$-effect. 
The fourth and final term can be interpreted as the time-averaged potential energy of the NIWs; indeed, using the vertical displacement in \eqn{displace1} and denoting averaging over the wave time scale $f_0^{-1}$ by $\av{ \cdot}$, we compute this as
\begin{equation}
\av { \int \frac{N^2 \zeta^2}{2}  \, \d \bx} = \frac{1}{4} \int N^2 | \nabla \chi|^2 \, \d \bx.
\end{equation}
{Here the left-hand side is the standard expression for the quadratic part of the potential energy in a Boussineq fluid in terms of vertical displacements \citep[e.g.][]{holl-mcin}.}
The total energy in the model could alternatively be defined as $\mc{H} + f_0 \mc{A}$. However, since $f_0 \mc{A} \gg \mathcal{H}$ is conserved independently, and $\mathcal{H}$ is the Noetherian conserved quantity associated with  time invariance (see \S\,\ref{sec:hamilton}), our separation appears more natural.

The energy and action are not the only conserved quantities for the coupled model. Clearly, the enstrophy and more generally the integrals
\begin{equation}
\int f(q) \, \d \bx
\end{equation}
of arbitrary functions $f$ of the PV are conserved, as in the standard quasi-geostrophic model. In fact, as we discuss in \S\,\ref{sec:hamilton}, the coupled model is Hamiltonian and additional conservation laws (e.g. linear and angular momentum) can be derived using Noether's theorem. 

\subsection{Scaling assumptions}

Our derivation of the coupled model relies on a number of approximations which we now detail. 
{
The parameters characterising the mean flow are the  Burger and Rossby numbers
\begin{equation}
\mathrm{Bu} = \frac{N^2H^2}{f_0^2L^2}  \quad \textrm{and} \quad \Ro = \frac{U_\mathrm{QG}}{f_0L},
\end{equation}
where $L$ and $H$ are the mean-flow horizontal and vertical scales, and $U_\mathrm{QG}$ is a typical mean velocity. These parameters are taken to satisfy $\textrm{Bu}=O(1)$ and $\Ro \ll 1$ in accordance with quasi-geostrophic theory.
The NIWs are characterised by two parameters analogous to $\mathrm{Bu}$ and $\Ro$, namely
\begin{equation}
\eps = \frac{N k}{f_0 m} \quad \textrm{and} \quad \alpha = \frac{U_\mathrm{NIW}}{f_0L}, \label{wburger}
\end{equation}
where $k$ and $m$ are typical horizontal and vertical wavenumbers, and $U_\mathrm{NIW}$ is a typical NIW horizontal velocity.
The parameter $\epsilon$, which measures the relative frequency shift of NIWs compared with $f_0$, is small: $\eps \ll 1$.
In the YBJ model \eqn{YBJ}, dispersion and mean-flow effects have similar orders of magnitudes provided that $\Ro = O(\eps^2)$, which we also assume. 
Note that this makes no specific assumption about the relative size of the  wave and mean horizontal scales which can be taken to satisfy $kL=O(1)$ (provided that $mH =O(\eps^{-1}) \gg 1$).

The parameter $\alpha$ controls the NIW amplitude. We choose its scaling relative to $\Ro$ in order that the NIW feedback affect{s} the mean motion at the same order as nonlinear vorticity advection. 
}
This imposes that
\begin{equation}
\Ro = O(\alpha^2), \quad{\textrm{hence}} \quad \alpha = O(\eps). \label{alphaeps}
\end{equation}
This scaling indicates that $U_\mathrm{NIW}/U_\mathrm{QG} = \alpha^{-1} \gg 1$, as is relevant to strong NIWs generated by intense storms  \citep[e.g.][]{DAsa1995}. It leads to a mean equation that is a modification of the quasi-geostrophic equation by wave effects. Had a smaller wave amplitude be assumed in order to model quieter conditions, say by taking $\Ro = O(\alpha)$, the wave effects would have been an $O(\Ro)$-factor smaller than advection in \eqn{MQGPV_Eq} and of comparable order as balanced corrections to quasi-geostrophy \citep[cf.][]{Zeitlin2003}. Because these corrections do not alter the qualitative properties of the quasi-geostrophic model, we prefer the scaling \eqn{alphaeps} to retain a model that is as simple as possible. {In spite of their relatively large amplitudes, the NIWs remain described by the YBJ equation which neglects all wave--wave interactions. This is justified on the ground that these interactions are remarkably weak for NIWs: first because NIWs triads cannot be resonant, and second because of a cancellation of the cubic terms associated with resonant quartets \citep{Falk1994,Zeitlin2003}.}

It is however important to note that our model is not fully  consistent from an asymptotic viewpoint.
The assumption of two different aspects ratios for  NIWs and  mean flow {--} implied by the condition $\eps \ll 1$ and $\textrm{Bu}=O(1)$ and best thought of as resulting from a disparity in vertical scales, $mH \gg 1$ {--} is not generally consistent. Indeed, small-scale NIWs generally lead to small-scale wave terms in \eqn{mqgpv} and hence to a pair $q$ and $\psi$ that varies on the wave scale (with a vertically-planar NIW field $\chi \propto \exp(\ii m z)$ a notable exception, see \S\,\ref{Sec_tw}). A consistent assumption would be to take $\mathrm{Bu} = O(\eps^2) \ll 1$. But this assumption is  less relevant to most of the ocean; it leads to a different balanced dynamics, namely frontal dynamics, with negligible wave--mean interactions \citep{Zeitlin2003}. 

While the model is heuristic, we regard it as valuable for its simplicity and because it respects key properties including conservation laws. The variational derivation of the wave--mean equations as detailed in the next sections makes this possible. This derivation starts with that of a coupled YBJ--primitive-equation (Eqs.\ (\ref{meanpri})--(\ref{wavefull}) below) which makes no assumption of quasi-geostrophy for the mean flow. This model, naturally more complex than \eqn{model}, is asymptotically consistent provided that $\alpha \ll \Ro^{1/2}/\epsilon^{1/2}$ so that the wave--wave interactions are negligible. It could serve as basis to obtain a balanced model for the mean flow that is more accurate than quasi-geostrophic and/or with relaxed assumptions on $\mathrm{Bu}$ so as to be fully consistent with the YBJ equation.

\section{Derivation of the coupled model} \label{sec:derivation}

We follow \citet{Salm2013} in deriving the Lagrangian-mean and wave equations from a variational formulation of the fluid equations rather than from the equations themselves. This is advantageous since it guarantees that the wave--mean model inherits conservation laws from the original hydrostatic--Boussinesq model. 
 While \citet{Salm2013} develops a general theory making no specific assumptions on the form of the perturbations to the mean flow, we focus on NIWs, assuming that the displacements $\bxi$ satisfy \eqn{displace1}. With this assumption, which can be viewed as a form of closure relying on a hypothesis of small wave amplitude, a natural step is to average the Lagrangian in the manner of \citet{Whit1974} to obtain a reduced Lagrangian that is a functional of the mean map $\bX$ and of the NIW amplitude $\chi$. This is described in \S\,\ref{Lagran}. Variations with respect to $\bs{X}$ (or rather its inverse) and $\chi$ are carried out in \S\,\ref{mean_dynamics} to obtain the mean and wave (YBJ) equations.

\subsection{Lagrangian and wave--mean decomposition} \label{Lagran}

The hydrostatic--Boussinesq equations \eqn{hydrobou} can be derived from the Lagrangian
\begin{eqnarray}
\mc{L}[\bx,p] = \int \left( \frac{1}{2}\left(\dot{x}^2+ \dot{y}^2\right)- \left( f_0 y+\frac{1}{2} \beta y^2 \right) \dot{x} + \theta  z + p \left( \left|\frac{\partial \bs{x}}{\partial \bs{a}} \right|-1 \right) \right) \, \d \bs{a}, \label{LagPhy}
\end{eqnarray}
where $\bs{a} = (a,b,\theta)$ are particle labels, with the (materially conserved) buoyancy $ \theta$ taken as third component, and $\bx(\bs{a},t)$ 
is the flow map \citep[e.g.][]{Salm2013}. The pressure $p(\bx,t)$ is 
a Lagrange multiplier enforcing the incompressibility constraint. Following standard GLM practice, we introduce the mean-map $\bX(\bs{a},t)$ and displacement $\bs{\xi}(\bx,t)$, with
\begin{equation}
\bx(\bs{a},t)=\bX(\bs{a},t) + \bs{\xi}(\bX(\bs{a},t){,t}).
\end{equation}
Following \citet{Salm2013}, we regard the Lagrangian as a functional of the inverse of the mean flow map, $\bs{a}(\bX,T)=\bX^{-1}(\bX,t)$, with $T=t$. Using the chain rule, \eqn{LagPhy} can be shown to take the form
\begin{eqnarray}
\mc{L}[\bs{a},\bs{\xi},p] &=& \int \left( J \left(\frac{1}{2}\br{\br{U+D_T\xi}^2+ \br{V+D_T\eta}^2}- \br{ f_0(Y+\eta)+\frac{1}{2} \beta (Y+\eta)^2 }  \right. \right. \nonumber\\
&& \qquad  \times \br{U+D_T\xi} \left. + \theta (Z+\zeta) \bigg) + p(\bs{X}) \br{ \left| \frac{\partial (\bX + \bs{\xi})}{\partial \bs{X}} \right|-J } \right) \, \d \bs{X}, \label{lagMea}
\end{eqnarray}
where $D_T = \partial_T + \bs{U} \cdot \nabla_3$, with $\bs{U}=\dot{\bX}=\bar{\bs{u}}^\mathrm{L}$ the Lagrangian-mean velocity and $\nabla_3$ the three-dimensional gradient with respect to $\bs{X}$, and $J = |\partial \bs{a}/\partial \bs{X}|$ is the Jacobian of the inverse mean map. 
In this expression, $\bs{U}$ should be thought as a differential function of $\bs{a}(\bX,T)$; an explicit form for it is obtained from the material invariance of the labels, $D_T \bs{a} =0$, as
\begin{eqnarray}
U = -\frac{1}{J} \jaco{a}{b}{\theta}{T}{Y}{Z},~ V = -\frac{1}{J} \jaco{a}{b}{\theta}{X}{T}{Z}, ~W = -\frac{1}{J} \jaco{a}{b}{\theta}{X}{Y}{T}. \label{velocity}
\end{eqnarray}

We next introduce the expansion
\begin{equation}
\bs{\xi} =  \bs{\xi}^{(1)}+  \bs{\xi}^{(2)}+\ldots,\label{ExpPer}
\end{equation}
of the NIW displacement into the Lagrangian \eqn{lagMea}, with $|\bs{\xi}^{(n)}| = O(\alpha^n)$. Retaining only terms in $\alpha^n,\, n\le 2$, which amounts to linearising the NIW dynamics, and averaging leads to the Lagrangian
\begin{eqnarray}
\av{\mc{L}} &=& \int \left(\frac{1}{2}J \br{ U^2 + V^2  + 2U D_T\av{\xi^{(2)}} + 2V D_T \av{\eta^{(2)}} + \av{\br{D_T\xi^{(1)}}^2} + \av{\br{D_T\eta^{(1)}}^2} } \right. \nonumber \\
  &&-J\br{ f_0Y+\frac{1}{2} \beta Y^2 } \br{ U + D_T\av{\xi^{(2)}}} - J\br{ f_0+\beta Y } \br{ \av{\eta^{(2)}}U + \av{\eta^{(1)} D_T \xi^{(1)}}} \nonumber \\
  &&\left. -J\frac{1}{2} \beta \av{\left({\eta^{(1)}}\right)^2} U + J\theta Z + J\theta \av{\zeta^{(2)}} \right.  \label{LA} \\
  &&\left. + P  \nabla_3 \cdot \left( \av {\bs{\xi}^{(2)}}  -\frac{1}{2}\av{ \bs{\xi}^{(1)}\cdot\nabla_3 \bs{\xi}^{(1)} }\right) + P(1-J) \right) \, \d\bs{X}, \nonumber
\end{eqnarray}
where
$\av{\cdot}$ denotes the average. It is standard in GLM theories that this average be defined as an arbitrary ensemble average. Here, a natural ensemble is that formed by a family of NIWs differing by a phase shift. Thus, an ensemble parameter $\gamma \in [0,2\pi]$ is introduced in (\ref{displace1}) to obtain the ensemble of leading-order wave fields
\begin{equation}
\xi^{(1)} + \ii \eta^{(1)} = \chi^{}_Z \, \ex^{-\ii (f_0 t + \gamma)} \quad \textrm{and} \quad \zeta^{(1)} = - \chi^{}_S \ex^{-\ii (f_0 t + \gamma)} + \cc \label{displace3}
\end{equation}
with $S=X + \ii Y$ and $\partial_{S}=(\partial_X-\ii\partial_Y)/2$. When there is a time-scale separation between the (fast) oscillation at frequency $f_0$ and the (slow) evolution of the amplitude $\chi$, averaging over $\gamma$ amounts to averaging over the fast time scale $f_0^{-1}$. Thus the ensemble average becomes physically relevant, and it leads to an averaged dynamics identical to that obtained by explicit perturbation expansions as demonstrated by \citet{Whit1974}. Note that our notation $\bxi^{(1)}(\bx,t)$ does not make the dependence of $\bxi^{(1)}$ on the ensemble parameter $\gamma$ explicit; our compact notation is justified by the fact that parameter $\gamma$ disappears completely from the problem after the (Whitham) average has been performed.
Note also that the truncation of the Lagrangian (\ref{LA}) to $O(\alpha)$ can be regarded as a closure in which the nonlinearity of wave dynamics is neglected.

To derive \eqn{LA}, we have used  that $\av{\bs{\xi}^{(1)}}=0$, that $\nabla_3 \cdot \bs{\xi}^{(1)}=0$ (stemming from the divergence-free property of NIWs), and that
\begin{equation}
 \left| \frac{\partial (\bX + \bs{\xi})}{\partial \bs{X}} \right|
 = 1 + \nabla_3 \cdot \bs{\xi}^{(2)} + \frac{1}{2} \nabla_3 \cdot \left( \bs{\xi}^{(1)} \nabla_3 \cdot \bs{\xi}^{(1)} - \bs{\xi}^{(1)} \cdot \nabla_3 \bs{\xi}^{(1)} \right) + O(\alpha^3), \label{jaco}
\end{equation}
as well as integration by parts. Importantly, we do not assume that $\av {\bs{\xi}^{(2)}}=0$ as is standard in GLM theory. Instead, we follow \citeauthor{Sowa2010}' (\citeyear{Sowa2010}) glm prescription which ensures that the mean motion is divergence free. As detailed in Appendix \ref{Aglm}, at the order we consider, this prescription amounts to taking
\begin{equation}
\av{\bs{\xi}^{(2)}} =  \frac{1}{2}  \av{\bs{\xi}^{(1)} \cdot \nabla_3 \bs{\xi}^{(1)}}. \label{glm}
\end{equation}
{Thus $\av{\bs{\xi}^{(2)}} \not= 0$ takes a value slaved to $\bs{\xi}^{(1)}$ (which contains terms in both $\ex^{\pm \ii (f_0 t + \gamma)}$) and hence to $\chi$.}
As \eqn{jaco} indicates, this ensures that the map $\bs{X} \mapsto \bs{X} + \bs{\xi}$ from mean to perturbed position is volume preserving: since the map $\bs{a} \mapsto \bs{X} + \bs{\xi}$ is volume preserving, this is also true for the map $\bs{a} \mapsto \bs{X}$, so the Lagrangian-mean velocity is divergence free. 

At this point, we can substitute the NIW-ansatz (\ref{displace3}), rewritten here as
into \eqn{LA} to obtain the averaged Lagrangian in terms of $\bs{a}$, $P$ and $\chi$. This leads to
\begin{eqnarray}
\av{\mc{L}} &=&  \int \left(\frac{1}{2}J (U^2 + V^2 ) - J\br{ f_0Y+\frac{1}{2} \beta Y^2 } U  + J\theta Z \right. \nonumber \\
&& \qquad + J\left( -\frac{\ii f_0}{4}(\chi^{}_ZD_T\chi_Z^*- \chi_Z^*D_T\chi^{}_Z) -\frac{1}{2}f_0\beta Y|\chi^{}_Z|^2 \right) \nonumber \\
&& \qquad  + J\br{-f_0YD_T\av{\xi^{(2)}} -f_0\av{\eta^{(2)}}U + \theta \av{\zeta^{(2)}} } + P (1-J) \bigg) \, \d \bs{X}. \label{MoLa}
\end{eqnarray}
To obtain this expression, we have retained only wave terms that are $O(1)$ or $O(\alpha^2/\Ro)$ relative to the size $U_\mathrm{QG}^2$ of the first term, assuming that $\beta L/f = O(\Ro)$ so that only a single wave term involving $\beta$ remains. Note that the linearisation of the NIW dynamics entailed by ignoring cubic terms in $\av{\mathcal{L}}$ {can be justified}: averaging eliminates cubic terms in $\bs{\xi}^{(1)}$, leaving cubic terms involving higher harmonics (with frequency $2f$), whose size can be estimated as $\epsilon \alpha^4/\Ro^2 = O(\alpha)$. The absence of resonant cubic terms has been noted by \citet{Falk1994} and \citet{Zeitlin2003} and is related to the possible elimination of advective nonlinearities by means of Lagrangian coordinates \citep{Falk1994,Hunt2013}.

The Lagrangian \eqn{MoLa} governs the  NIW--mean flow system: when \eqn{displace3} and \eqn{glm} are used to express $\bs{\xi}^{(2)}$ explicitly as
\begin{subequations} \label{xi2}
\begin{align}
\av{ \xi^{(2)} }  &= \frac{1}{4} ( \chi^{}_Z \chi^*_{ZS} - \chi^{}_{S}\chi^*_{ZZ}  ) + \mathrm{c.c.}, \\
\av{ \eta^{(2)} } &= \frac{\ii}{4} ( \chi^{}_Z \chi^*_{ZS} - \chi^{}_{S}\chi^*_{ZZ}  ) + \mathrm{c.c.}, \\
\av{ \zeta^{(2)} } &= \frac{1}{2} ( - \chi^{}_Z \chi^*_{SS^*} + \chi^{}_{S}\chi^*_{ZS^*} )  + \mathrm{c.c.},
\end{align}
\end{subequations}
$\mathcal{L}$ is a functional of $\bs{a}$, $\chi$ and $P$ from which primitive equations for the mean flow coupled to a YBJ-like equation for the NIWs can be derived systematically. This is carried out in the next subsection, \S\,\ref{mean_dynamics}. The reduced quasi-geostrophic model \eqn{model} is then derived in \S\,\ref{sec:qg}.

\subsection{Coupled YBJ--primitive-equation model}  \label{mean_dynamics}

Taking the variation $\delta P$ of the action $\int \av{\mc{L}} \, \d t $ with the Lagrangian (\ref{MoLa}) and using  (\ref{xi2}) we obtain 
\begin{eqnarray}
J=1,
\end{eqnarray}
confirming that the mean map is volume preserving. Thus the Lagrangian-mean velocity is divergence free:
\begin{equation} 
\nabla_3 \cdot \bs{U}=0. \label{MeanInc2}
\end{equation}

The mean equations of motion can now obtained from the stationarity of $\int \av{\mathcal{L}} \, \d t$ with respect to variations $\delta \bs{a}$. It is convenient to use the energy-momentum formalism as proposed by Salmon (2013). Computations detailed in Appendix \ref{Mean_Mom} lead to the momentum equations in the form
\begin{subequations} \label{meanpri}
\begin{align}
D_T U - (f_0 + \beta Y) V + \partial_X P =& \frac{\ii f_0}{2} \left( D_T \chi^{}_Z \chi_{XZ}^* - D_T  \chi_Z^* \chi^{}_{XZ}  \right) - \frac{1}{2}f_0\beta \partial_X (Y |\chi^{}_Z|^2) \nonumber \\
& + f_0 \av{ D_T \eta^{(2)} - U \eta^{(2)}_X + V \xi^{(2)}_X} + \theta \av{\zeta^{(2)}_X} , \label{meanpri1} \\
D_T V + (f_0 + \beta Y) U + \partial_Y P =& \frac{\ii f_0}{2} \left( D_T  \chi^{}_Z \chi_{YZ}^* - D_T \chi_Z^* \chi^{}_{YZ}  \right) - \frac{1}{2}f_0\beta \partial_Y (Y|\chi^{}_Z|^2)  \nonumber \\
& + f_0 \av{ - D_T \xi^{(2)} - U \eta^{(2)}_Y + V \xi^{(2)}_Y} + \theta \av{\zeta^{(2)}_Y} , \label{meanpri2} \\
- \theta + \partial_Z P =& \frac{\ii f_0}{2} \left( D_T \chi^{}_Z \chi_{ZZ}^* - D_T \chi_Z^* \chi^{}_{ZZ} \right) - \frac{1}{2}f_0\beta \partial_Z(Y |\chi^{}_Z|^2) \nonumber \\
& + f_0 \av{  - U \eta^{(2)}_Z + V \xi^{(2)}_Z} + \theta \av{\zeta^{(2)}_Z} . \label{meanpri3}
\end{align}
\end{subequations}
These are completed by the buoyancy equation
\begin{equation}
D_T \theta = 0 \label{MeanBuo}
\end{equation}
which expresses that $\theta$ is a label. The left-hand sides of Eqs.\ \eqn{MeanInc2}--\eqn{MeanBuo} recover the hydrostatic--Boussinesq equations \eqn{hydrobou} for the mean flow; the right-hand sides, which can be written completely in terms of $\chi$, describe the impact of the NIWs on the mean flow.


Taking the variation $\delta \chi^*$  of the Lagrangian \eqn{MoLa} after using (\ref{xi2}) for $\bs{\xi}^{(2)}$ leads to the wave equation
\begin{eqnarray}
(D_T \chi^{}_Z)^{}_Z - \ii \beta Y \chi^{}_{ZZ} + \frac{\ii}{2} \left( (V\chi^{}_Z)^{}_{ZS} - (V\chi^{}_{S})^{}_{ZZ} - (V\chi^{}_{ZS*})^{}_Z + (V\chi^{}_{ZZ})^{}_{S*} \right) & \nonumber \\
+\frac{1}{2} \left( (U\chi^{}_Z)^{}_{ZS} - (U\chi^{}_{S})^{}_{ZZ} - (U\chi^{}_{ZS^*})^{}_Z - (U\chi^{}_{ZZ})^{}_{S^*} \right) & \label{wavefull} \\
+\frac{\ii}{f_0}\left( -(\theta \chi^{}_Z)^{}_{SS^*} + (\theta \chi^{}_{S})^{}_{ZS^*} + (\theta \chi^{}_{SS^*})^{}_Z - (\theta \chi^{}_{ZS})^{}_{S^*} \right) &= 0.  \nonumber   
\end{eqnarray}
This equation can be interpreted as a generalisation of the YBJ equations which makes no assumption that the mean flow is quasi-geostrophic or steady. 

Together, Eqs.\ (\ref{meanpri}), (\ref{MeanBuo}) and (\ref{wavefull}) constitute a closed model for the joint evolution of the wave and the mean flow. This model is complex and we prefer to focus our analysis on its quasi-geostrophic approximation introduced in \S\,\ref{sec:coupledmodel} and derived in the next subsection. It is nonetheless worth noting that the full model has two simple conservation laws.
%
%
%
The first is obtained by multiplying \eqn{wavefull} by $\chi^*$ and adding the complex conjugate of the resulting equation. Integrating over space and making liberal use of integration by  parts yields the wave-action conservation
\begin{equation}
\frac{\d}{\d t} \int |\chi^{}_Z|^2 \, \d \bX = 0. \label{PEaction}
\end{equation} 
This conservation law is associated with the obvious symmetry $\chi^{} \mapsto \ex^{\ii \gamma} \chi^{}, \, \gamma \in \mathbb{R}$, of the Lagrangian (\ref{MoLa}) and can therefore also be obtained from Noether's theorem \citep[e.g.][]{gold80} in the form
\begin{equation}
\frac{\d}{\d t} \int \left(\ii \chi^{} \frac{\delta}{\delta \chi^{}_{T}} - \ii \chi^* \frac{\delta}{\delta \chi^*_{T}}\right) \av{\mathcal{L}} \, \d \bX = 0,
\end{equation}
thus justifying the terminology of action.
The second conservation law is that of energy. It is best obtained from the Lagrangian \eqn{MoLa}.  
The general form of the conserved energy, {associated with the symmetry $T \mapsto T+\delta T$, also follows from Noether's theorem. This yields the energy in the form} 
\begin{eqnarray}
\int \left( a_T^i \frac{\delta}{\delta a_T^i} + \chi^{}_T \frac{\delta}{\delta \chi^{}_T} + \chi_T^* \frac{\delta}{\delta \chi_T^*} - 1  \right) \av{\mathcal{L}} \,  \d \bX,
\end{eqnarray}
which implies that the energy is readily deduced from $\mathcal{L}$ using the following rules: 
terms that are quadratic in $\bs{U}$ (and hence in $a^i_T$) or $\chi^{}_T$ are retained,  terms that are linear are omitted, and terms that contain no time derivatives change sign. So the energy conservation reads
\begin{eqnarray}
\frac{\d}{\d t} \int\br{ \frac{1}{2} (U^2 + V^2 )  - \theta (Z + \av{\zeta^{(2)}} ) + \frac{1}{2}f_0\beta Y|\chi^{}_Z|^2}  \d \bX = 0 \label{energy_pri}
\end{eqnarray}   
using that $J=1$.
This is a remarkably simple expression in which the effect of the waves arises only through the potential-energy term $- \theta \av{\zeta^{(2)}}$ and the $\beta$-term. Surprisingly perhaps, it is simpler than the analogous energy that is conserved in the (uncoupled) YBJ model \citep{Vann2014}. 

%
%

\subsection{Quasi-geostrophic approximation} \label{sec:qg}

We now derive an approximation to the mean and wave equations in the quasi-geostrophic limit $\Ro \to 0$. The standard quasi-geostrophic model cannot be derived in a simple manner from the variational formulation of the primitive equations \citep[see][however]{Bokh1998,Oliv2006}, and the same difficulty arises here. We therefore derive the quasi-geostrophic approximation of the mean equations directly from the momentum equations \eqn{meanpri}, retaining a variational argument for the wave part only. That the approximations made in both parts of the model are consistent is confirmed by the fact that the resulting coupled model has a Hamiltonian structure, as discussed in \S\,\ref{sec:hamilton}.

In the quasi-geostrophic approximation, the buoyancy is decomposed into a $Z$-dependent mean part and a perturbation according to
\begin{equation}
\theta = \bar \theta(Z) + \theta' = \int^Z N^2(z) \, \d z + \theta'. \label{theta}
\end{equation}
To leading order in $\Ro$, the mean equations \eqn{meanpri} then reduce to
\begin{eqnarray}
f_0 V &=& \partial_X \left( P - \bar \theta \av{ \zeta^{(2)}} \right), \label{geo1} \\
-f_0 U &=& \partial_Y \left( P - \bar \theta \av{ \zeta^{(2)}} \right), \label{geo2} \\
\theta' &=& \partial_Z  \left( P - \bar \theta \av{ \zeta^{(2)}} - \int^Z \d z  \int^{z} N^2(z')\, \d z' \right)  + N^2 \av{ \zeta^{(2)}},\label{geo3}
\end{eqnarray}
and are recognised as expressing geostrophic and hydrostatic balance.
This leads to the introduction of a streamfunction $\psi$ such that
\begin{equation}
U = - \psi_Y, \ \ V = \psi_X  \ \ \textrm{and} \ \ \theta' = f_0 \psi_Z + N^2 \av{ \zeta^{(2)}}.
\end{equation}
Using this, the buoyancy conservation becomes
\begin{equation}
D^0_T  \left( f_0 \psi_Z + N^2 \av{ \zeta^{(2)}} \right) + N^2 W  = 0, \label{buo3}
\end{equation}
where $D_T^0 = \partial_T + \partial(\psi,\cdot)$. 

A closed equation for $\psi$ can now be derived from \eqn{meanpri} and \eqn{buo3} in a familiar way: taking the horizontal curl of \eqn{meanpri1}--\eqn{meanpri2} and keeping terms up to $O(U^2/L^2)$ we obtain
\begin{eqnarray}
D_T^0 \br{\beta Y + V_X-U_Y + \frac{\ii f}{2}\partial(\chi^{}_Z,\chi_Z^*) + f_0(\av{\xi^{(2)}_X}+\av{\eta^{(2)}_Y})} - f_0W_Z = 0.
\end{eqnarray}
Substituting \eqn{buo3} to eliminate $W$ leads to the conservation equation
\begin{equation}
D_T^0 q = 0, \ \ \textrm{where} \ \  q = \beta {Y} + \Delta \psi + \frac{\ii f_0}{2} \partial(\chi_Z^*,\chi^{}_Z) + f_0 \nabla_3 \cdot \av{\bs{\xi}^{(2)}}, \label{QGPV0}
\end{equation} 
{with $\Delta$ defined in (\ref{Delta}),} is the QGPV.
A direct computation using (\ref{xi2}) gives the last term explicitly as
\begin{equation}
\nabla_3 \cdot \av{\bs{\xi}^{(2)}} = G(\chi^*,\chi), \label{divxi2}
\end{equation}
with the symmetric bilinear operator $G$ defined in \eqn{G}. Replacing $\bs{X}$ by $\bs{x}$ as independent variable reduces the QGPV equation \eqn{QGPV0} to the form announced in \eqn{MQGPV_Eq}. An alternative derivation based on  potential-vorticity conservation and valid for an arbitrary definition of the Lagrangian average 
is presented in Appendix \ref{app:altder}.
The vertical boundary conditions \eqn{Boun_Dyn} associated with the QGPV equation are derived by applying the no-normal-flow condition $W=0$ at $z=z^\pm$ to \eqn{buo3} and noting from \eqn{glm} that $\av {\zeta^{(2)}}=0$ at $z^\pm$ follows from the fact that $\zeta^{(1)}=0$ there.

The NIW equation associated with \eqn{QGPV0} is best derived by introducing the geostrophic and hydrostatic conditions into the averaged Lagrangian \eqn{MoLa} then taking variations with respect to $\chi$ or $\chi^*$. The wave part of the Lagrangian is readily found from \eqn{MoLa} to be
\begin{eqnarray}
\av{\mathcal{L}}_\mathrm{NIW} &=& \int \Bigg( -\frac{\ii f_0}{4}(\chi^{}_Z D^0_T\chi_Z^*- \chi_Z^*D^0 _T\chi^{}_Z) -\frac{1}{2}f_0\beta Y|\chi^{}_Z|^2  \nonumber \\
&& \qquad \qquad \left. - f_0 \psi  \nabla_3 \cdot \av{\bs{\xi}^{(2)}} + \int^Z N^2(z) \, \d z \, \av{\zeta^{(2)}} \right)
 \, \d \bs{X}, \label{lagniw}
\end{eqnarray}
where we have used that $J=1$, integration by parts, and neglected a term in $\av{\zeta^{(2)}}^2$ The terms depending on $\bs{\xi}^{(2)}$ can now be written in terms of $\chi$ using \eqn{divxi2} and the observation that 
\begin{equation}
\av{\zeta^{(2)}} = \frac{1}{2} \partial_Z \av{\left(\zeta^{(1)}\right)^2} + \cdots = \frac{1}{4} \partial_Z |\nabla \chi|^2 + \cdots, \label{useful}
\end{equation}
where $\cdots$ denotes the horizontal divergence of an irrelevant vector. This simplifies \eqn{lagniw} into
\begin{eqnarray}
\av{\mathcal{L}}_\mathrm{NIW} &=& - \int \left( \frac{\ii f_0}{4}(\chi^{}_Z D^0_T\chi_Z^*- \chi_Z^*D^0 _T\chi^{}_Z) +\frac{1}{2}f_0\beta Y|\chi^{}_Z|^2 \right. \nonumber \\
&& \qquad \qquad \left.   + f_0 \psi  G(\chi^*,\chi) + \frac{1}{4} N^2 |\nabla \chi^{}|^2 \right)
 \, \d \bs{X}. \label{oho}
\end{eqnarray}
To take the variations of the corresponding action, it is convenient to introduce the symmetric bilinear operator $\hat G$ dual to $G$ in the sense that
\begin{equation}
\int \psi G(\chi^*,\chi) \, \d \bX = \int \chi^* \hat{G}(\psi,\chi) \, \d \bX. \label{GhatG}
\end{equation}
The variation $\delta \chi^*$ then gives
\begin{equation}
(D^0_T \chi)^{}_Z + \ii \beta Y \chi^{}_{ZZ} + \frac{\ii N^2}{2f_0} \nabla^2 \chi^{} - 2 \ii \hat{G}(\psi,\chi^{}) = 0. \label{YBJ-1}
\end{equation}
From its definition and (\ref{G}) $\hat G(\psi,\chi)$ is calculated to be
\begin{equation}
\hat G(\psi,\chi) = \frac{1}{4}\br{ 2\nabla \psi_Z \cdot \nabla \chi^{}_Z - \nabla^2\psi \chi^{}_{ZZ} - \psi_{ZZ}\nabla^2\chi} \label{ggg}
\end{equation}
and is recognised as the negative of YBJ's bracket $[[\cdot,\cdot]]$. Introducing \eqn{ggg} into \eqn{YBJ-1}, dropping the superscript $0$ from $D_T^0$ and replacing $\bX$ by $\bx$ leads to the YBJ equation in the form \eqn{YBJ}.

\section{Conservation laws and Hamiltonian structure} \label{sec:hamilton}

We now derive conservation laws satisfied by the coupled model \eqn{model}. We start by the conservation law identified in YBJ: multiplying \eqn{YBJ} by $\chi^*$ and integrating yields
\begin{equation}
\int \left( - \chi_z^* \partial_t \chi^{}_z + \psi \partial(\chi_z^*,\chi^{}_z) - \ii \beta y |\chi^{}_z|^2 - \frac{\ii N^2}{2f_0} |\nabla \chi^{} |^2 - 2 \ii \psi G(\chi^*,\chi) \right) \, \d \bx =0,
\end{equation}
after using integration by parts. Adding the complex conjugate and using the symmetry of $G$ and antisymmetry of  $\partial(\cdot,\cdot)$ gives
\begin{equation}
\frac{\d}{\d t} \int |\chi^{}_z|^2 \, \d \bx = 0. \label{WaveAc}
\end{equation}
Thus, the wave action $\mathcal{A}$ defined in \eqn{Cons_Wave_Act} is conserved. This conservation law is identical to that obtained for the YBJ--primitive-equation model in (\ref{PEaction}) and, as checked below using the Hamiltonian structure of the YBJ-QG model, also associated with an invariance with respect to phase shifts of the amplitude $\chi$.

Next we derive an energy conservation law. Multiplying the QGPV equation \eqn{MQGPV_Eq} by $\psi$, integrating and using the definition \eqn{mqgpv} of $q$ gives
\begin{eqnarray}
&& \int \left( \frac{1}{2} \partial_t \left({|\nabla \psi|^2} + \frac{f_0^2}{N^2} \psi_z^2 \right)- \frac{\ii f_0 \psi}{2} \left(  \partial(\chi_{zt}^*,\chi^{}_z) + \partial(\chi_{z}^*,\chi^{}_{zt}) \right) \right. \nonumber \\
&& \qquad \qquad \qquad  \qquad \qquad \qquad  - f_0 \psi \left( G(\chi_t^*,\chi) + G(\chi^*,\chi^{}_t) \right) \bigg) \, \d \bx = 0. \label{en1}
\end{eqnarray}
Multiplying the YBJ equation \eqn{YBJ} by $\ii f_0 \partial_t \chi^*/2$, integrating and adding the complex conjugate gives
\begin{eqnarray}
&& \int \left( \frac{\ii f_0 \psi}{2} \left( \partial(\chi_{zt}^*,\chi^{}_z)+\partial(\chi_{z}^*,\chi^{}_{zt}) \right) + \frac{f_0 \beta y}{2}\partial_t |\chi^{}_z|^2  \right. \nonumber \\ 
&& \qquad \qquad \qquad \left. + \frac{N^2}{4} \partial_t |\nabla \chi|^2 + f_0 \psi \left( G(\chi_t^*,\chi) + G(\chi^*,\chi^{}_t) \right) \right) \, \d \bx = 0, \label{en2}
\end{eqnarray}
where the relation \eqn{GhatG} between $G$ and $\hat G$ is used. Adding \eqn{en1} and \eqn{en2} leads to
\begin{equation}
\frac{\d}{\d t} \int \frac{1}{2} \left(|\nabla \psi|^2 + \frac{f_0^2}{N^2} \psi_z^2 + {f_0}\beta y |\chi^{}_z|^2 + \frac{1}{2} N^2 |\nabla \chi|^2  \right) \, \d \bx = 0,
\end{equation}
and hence to the conservation of the energy $\mathcal{H}$ in \eqn{TotEne}. This energy conservation can be recognised as the QG approximation of primitive-equation energy (\ref{energy_pri}): the first two terms are the usual QG approximation of the mean kinetic and potential energy; the third term is unchanged; the fourth term is an approximation to $\theta \av{\zeta^{(2)}}$ obtained by noting that $\theta \approx \int^z N^2(z') \, \d z'$ and using (\ref{useful}). 

The coupled model \eqn{model} is in fact Hamiltonian. The Hamiltonian structure \citep[e.g.][]{shep90}, which can be obtained by inspection, is conveniently written using the amplitude of the horizontal NIW displacement $\phi=\chi^{}_z$, its complex conjugate $\phi^*$, $q$ and $\theta^\pm$ as dynamical variables. Grouping these in a vector $\bs{\phi}$,
it can be checked that the governing equations \eqn{model} are recovered from
\begin{equation}
\bs{\phi}_t = \mathcal{J} \frac{\delta \mathcal{H}}{\delta \bs{\phi}}, 
\end{equation}
where
\begin{equation}\label{Ham_Str}
\mathcal{J} = 
\begin{pmatrix}
 0 & -2\ii / f_0 & 0 & 0 & 0 \\
2 \ii /f_0 & 0 &  0 & 0 & 0 \\ 
0 & 0 & -\partial(q,\cdot) & 0 & 0 \\
0 & 0 & 0 & (N^+)^2 f_0^{-1} \partial(\theta^+,\cdot) & 0 \\
0 & 0 & 0 & 0 & -(N^-)^2 f_0^{-1}\partial(\theta^-,\cdot)
\end{pmatrix}
\end{equation}
and the Hamiltonian is
\begin{equation}
\mc{H} = \frac{1}{2} \int \left( |\nabla\psi|^2 + \frac{f_0^2}{N^2}|\psi_z|^2 + f_0\beta y |\phi|^2 + \frac{N^2}{2}  \left| \nabla \int^{z}  \phi({\tilde{z}}) \, \d {\tilde{z}} \right|^2 \right) \, \d \bs{x}.
\end{equation}
The streamfunction $\psi$ is here regarded as a functional of {$q$ and} $\bs{\phi}$ defined by
\begin{equation}
\psi = \Delta^{-1} \left( q - \beta y - \frac{\ii f_0}{2} \partial(\phi^*,\phi) - f_0 G\left(\int^{z}  \phi({\tilde{z}})^* \, \d {\tilde{z}},\int^{z}  \phi({\tilde{z}}) \, \d {\tilde{z}} \right) \right)  \label{Stream_Fun}
\end{equation}
with 
\begin{equation}
\left. \psi_z \right|_{z=z^\pm} = f_0^{-1} \theta^\pm
\end{equation}
{ following (\ref{Boun_Dyn}).}

The Hamiltonian structure provides a systematic route to the derivation of conservation laws using Noether's theorem. We note that the Hamiltonian flow associated with the wave action $\mathcal{A} = f_0 \int |\phi|^2 \, \d \bx/2$, { namely $\mathcal{J}\delta \mathcal{A}/\delta \bs{\phi}$, is $(-\ii \phi, \ii \phi^*,0,0,0)^T$}. This is recognised as the generator of the continuous transformation $\phi \mapsto \phi \exp(-\ii \gamma), \, \gamma \in \mathbb{R}$, an obvious symmetry of $\mathcal{H}$.
The invariance of $\mathcal{H}$ with respect to translations and horizontal rotations gives rise to conserved linear and angular momenta. For instance, the conserved $x$-momentum is readily shown to be 
\begin{equation}
\begin{aligned}
\mathcal{M}_x &= \int \left( \frac{\ii f_0}{4} \left(\phi^* \phi_x - \phi^*_x \phi\right) - q y \right) \, \d \bx + f_0 \int \left((N^+)^{-2} \theta^+ - (N^-)^{-2} \theta_- \right) y \, \d x \d y \\
&= \int U  \, \d \bx.
\end{aligned}
\end{equation}
Additional conserved quantities are of course the same Casimir invariants as in three-dimensional quasi-geostrophic dynamics, namely the volume integrals of arbitrary functions of $q$ and surface integrals of arbitrary functions of $\theta^\pm$ \citep{shep90}.

\section{Implications} \label{sec:implications}

We now  discuss some implications of the conservation of energy (\ref{TotEne}) and action (\ref{Cons_Wave_Act}) for ocean dynamics. First, we note that the action conservation implies that the NIW amplitude remains zero if it is initially so: thus spontaneous generation of NIWs is impossible in this model, unsurprisingly since it is expected to be exponentially small in  $\Ro$  \citep{Vann2013} and thus much smaller than neglected terms. {Second, the energy conservation indicates that the decrease in NIW scales induced by the $\beta$-effect in the absence of a flow, $\psi=0$, is necessarily accompanied by an equatorward drift of the NIWs, consistent with WKB results \citep{Garr2001}.}  

A third, more striking, conclusion is that conservation laws show unambiguously that oceanic NIWs forced by atmospheric winds provide an energy sink for the mean flow. To see how, consider NIWs forced at some initial time $t=0$ with horizontal scales large enough that $\chi^{}_0=\chi^{}(t=0)$ has negligible horizontal gradient i.e. $\nabla \chi^{}_0 \approx 0$. This is a reasonable approximation since
NIWs are generated by atmospheric storms whose scales are ten or more times the scale of oceanic eddies. 
Initially, NIWs make no contribution to the  energy $\mathcal{H}$ which then purely consists of the mean-flow energy. As time progresses, the advection and refraction of the waves by the mean flow lead to a scalar cascade in the NIW field, 
producing horizontal scales similar to, or smaller than, the eddy scale. As a result, $|\nabla \chi|$ grows since $|\chi|$ is constrained by wave-action conservation. According to (\ref{TotEne}), the contribution of $|\nabla \chi|^2$ to the energy must be balanced by a decrease in the energy of the mean flow. Physically, the mechanism for this energy exchange is clear: as the horizontal scale of the NIWs decreases, their potential energy increases, necessarily at the expense of the mean energy since the NIW kinetic energy $f_0 \mathcal{A}$ is conserved. This mechanism can be suggestively termed `stimulated wave generation' to distinguish it from spontaneous generation (ruled out in our model) and complete an electromagnetic analogy \citep[e.g.][]{quantum}.

The explicit form of (\ref{TotEne}) and (\ref{Cons_Wave_Act}) enables us to make quantitative predictions. Suppose that the NIWs initially have a typical vertical scale $m_0^{-1}$, corresponding for example to the depth of the mixed layer. Suppose too that at some final time $t$, the various processes governing their dynamics have led to typical horizontal and vertical scales $k^{-1}$ and $m^{-1}$ and to typical amplitudes $|\chi|$. The conservation of wave action (\ref{Cons_Wave_Act}) implies that
\begin{equation}
 \frac{f_0 m_0^2}{2} |\chi^{}_0|^2 \approx \frac{f_0 m^2}{2} |\chi|^2.
\end{equation}
Correspondingly, the kinetic energy of the NIW per unit volume, $\mathcal{K}_\mathrm{NIW} \approx f_0^2 m^2 |\chi|^2/2$ remains unchanged. The potential energy, on the other hand, increases from $0$ to $\mathcal{P}_\mathrm{NIW}  \approx N^2 k^2 |\chi|^2/4$. We therefore conclude that the NIWs extracts from the mean-flow an energy \begin{equation} \label{aa}
- \mathcal{E}_\textrm{QG} =  \mathcal{P}_\mathrm{NIW} = \frac{N^2 k^2}{2 f_0^2 m^2} \mathcal{K}_\mathrm{NIW} =  \frac{\eps^2}{2} \mathcal{K}_\mathrm{NIW}  
\end{equation}
per unit volume. {Because the dispersion relation of NIWs is $\omega = f_0 (1 + \eps^2/2)$ 
(as follows from the dispersion term in (\ref{YBJ}) or from a Taylor expansion of the inertia-gravity-wave frequency $\omega = (f_0 + N^2 k^2/m^2)^{1/2}$),} $\eps^2/2$ can also be rewritten as $\Delta \omega/f_0$, the relative frequency shift away from $f_0$.

Since one of the main open questions in ocean dynamics concerns the dissipation of mesoscale energy, it is natural to ask whether the mechanism we have identified could be a significant contributor. Assuming that the process of NIW generation followed by their cascade to small scale occurs in a continuous fashion, (\ref{aa}) can be turned into an expression for the power rate extracted from the mean flow,
\begin{equation}
-  \dot{\mathcal{E}}_\textrm{QG} = \frac{\eps^2}{2} \dot{\mathcal{K}}_\mathrm{NIW},
\end{equation} 
where $\dot{\mathcal{K}}_\mathrm{NIW}$ is the power  injected into NIWs by winds. Integrating over the whole ocean, this power  is estimated as $0.6$ TW in \citet{Wuns2004}.
It is unclear what a realistic value of $\eps^2/2$ might be: if we take $k$ and $m$ as representative of typical NIWs, $\eps^2/2 = \Delta \omega / f_0$ can be interpreted as the width of the inertial peak relative to $f_0$, and a value of $\eps^2/2=0.2$ is plausible. This leads to a sink of $0.12$ TW, comparable, for instance, with the  $0.1$ TW estimated for the dissipation caused by bottom drag \citep{Wuns2004}. There is considerable uncertainty in these estimates however, in particular because it is not clear what the final values of $k$ and $m$ ought to be and whether the impact of NIWs is restricted to the upper parts of the ocean. Furthermore, the scale cascade can be expected to lead to values of $\eps^2/2$ that are not small, e.g.\ through the mechanism of wave capture \citep{Badu1993,Buhl2005} which suggests that $\eps$ stabilises at $O(1)$ values. While our model ceases to be valid then -- and the crucial feature of conserved wave kinetic energy ceases to hold -- one can expect energy to be transferred from mean flow to the waves throughout the cascading process. Our argument above, necessarily limited to $\eps \ll 1$, may therefore underestimate the amount of energy extracted from the mean flow. It would certainly be valuable to test the efficiency of the process through detailed numerical simulations.  


\section{Two-dimensional models} \label{sec:2d}

In this section we discuss two two-dimensional models that are deduced from the YBJ-QG model under certain symmetry assumptions. These models are useful to study the NIW-mean interactions in a simplified context.  

\subsection{Slice model}

Neglecting the $\beta$-effect, we consider solutions that are independent of $y$. This reduces (\ref{model}) to
\begin{subequations} \label{cc}
\begin{align}
\chi^{}_{zzt} + \frac{\ii N^2}{2f_0}\chi^{}_{xx} + \frac{\ii}{2}(\psi_{xx}\chi^{}_{zz}+\psi_{zz}\chi^{}_{xx}-2\psi_{xz}\chi^{}_{xz}) &= 0, \label{ab} \\
\partial_t \br{ \psi_{xx} + \partial_z\br{\frac{f_0^2}{N^2}\psi_z} + \frac{f_0}{4}(2|\chi^{}_{xz}|^2 - \chi^{}_{zz}\chi^*_{xx}-\chi^*_{zz}\chi^{}_{xx}) }&=0. \label{bb}
\end{align}
\end{subequations}
Because advection disappears, (\ref{bb}) can be integrated in time to provide the streamfunction in terms of $\chi$, leaving (\ref{ab}) as the sole prognostic equation. 

We illustrate the interest of this model by presenting the result of a numerical simulation examining the impact of NIWs on a barotropic mean flow using a setup based on that of {\citet{Balm-et-al1998}}.
In this setup, NIWs initialised near the surface propagate vertically as a result of their interactions with the one-dimensional mean flow 
\begin{equation}
\nabla^\bot \psi =  (0,U_\mathrm{QG} \sin(2 \pi x / L)),
\end{equation}
where $L$ is the length of the domain. The coupled model enables us to study the feedback of the NIWs on this mean flow. 

We carried out simulations using a pseudospectral implementation of (\ref{cc}), with a domain $(x,z) \in [0,L] \times [-H,0]$ where $L = 80$ km and $H=4\,200$ m.   The Coriolis frequency is taken as $f_0= 10^{-4}\, \mr{s^{-1}}$ and a constant Brunt--V\"ais\"al\"a frequency $N=8\times 10^{-3} \,\mr{s^{-1}}$, somewhat smaller than that in \citet{Balm-et-al1998}, is used. 
The maximum mean velocity is $U_{\mathrm{QG}} = 0.08 \,\mr{m\,s^{-1}}$. The NIWs are initially confined within the mixed layer with a characteristic depth $H_\mathrm{m}=50 \,\mr{m}$, with the form ${\chi}_{0z} = U_{\mr{NIW}}\exp(-(z/H_m)^2)$ where $U_{\mr{NIW}}=0.8 \,\mr{m\,s^{-1}}$ . The corresponding dimensionless parameters are $\Ro = 0.01$, $\alpha = 0.1$ and $\epsilon = 0.05$,  so $\Ro^{1/2} = \alpha \approx \epsilon$, consistent with our scaling assumptions.

\begin{figure}
\centering
\includegraphics[width=7.5cm]{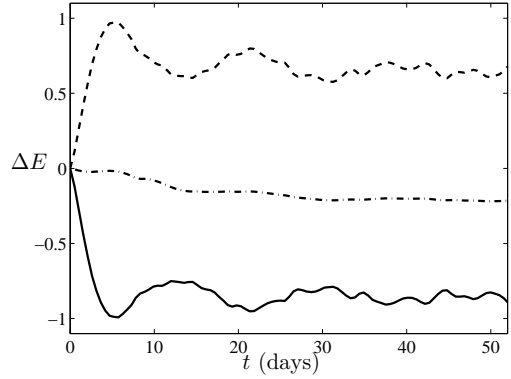}
\caption{Energy exchange in the slice model: the {changes in the} mean energy (solid line), NIW energy (dashed line) and total energy (dotted line) are shown as functions of time. 
{ These energy changes are normalized by the initial mean flow energy in the mixed layer, $z \in [-50,0]$ m.}
The increase of NIW (potential) energy is offset by a mean energy loss, resulting in a total energy that is conserved up to a small hyperviscous dissipation added for numerical stability.}
\label{xzenergy}
\end{figure}

\begin{figure}
\centering
\hspace*{-1.5cm}\includegraphics[width=1.2\textwidth]{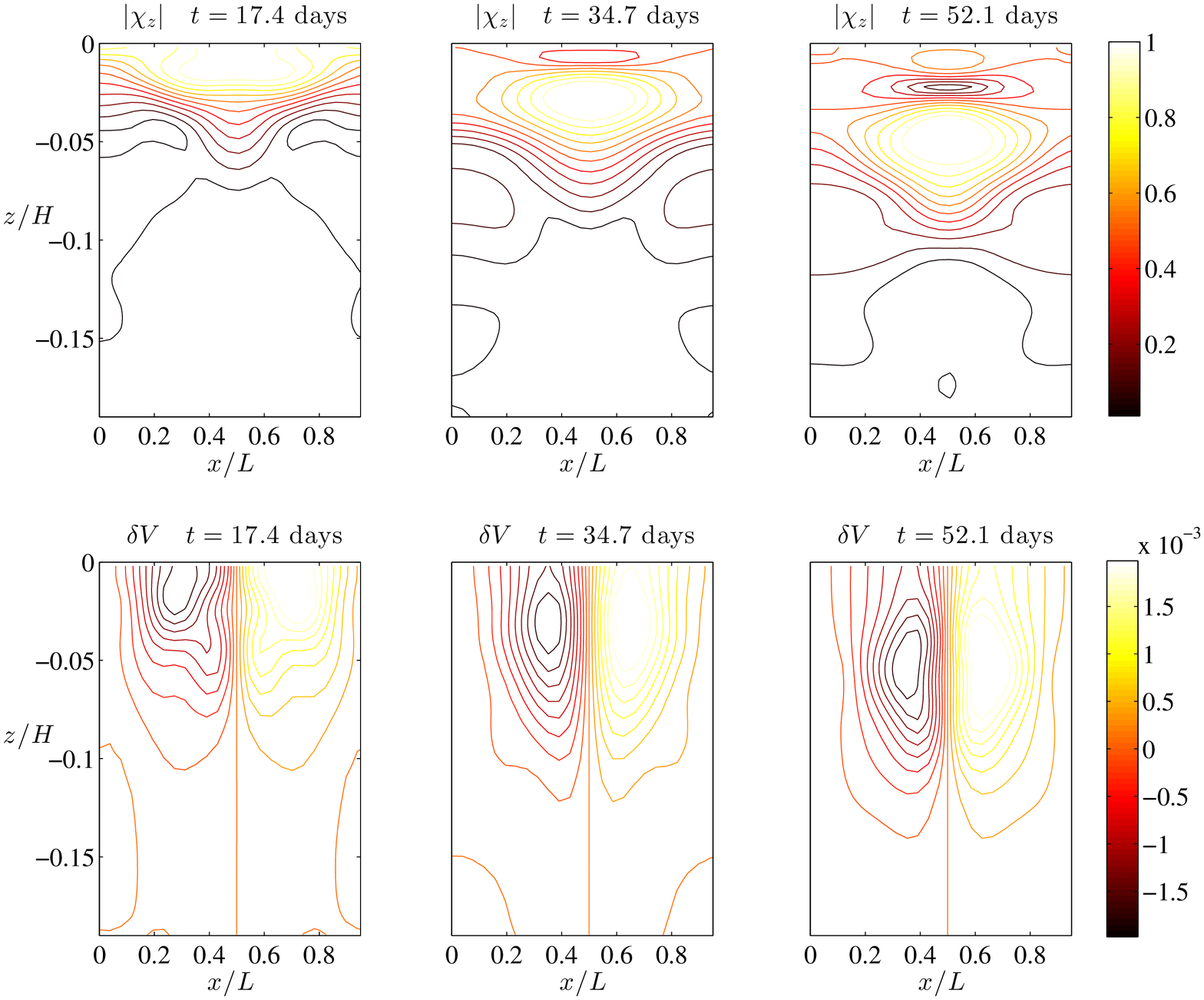}
\caption{Wave amplitude $|\chi^{}_z|$ (upper panels) and change in the mean velocity $V=\psi_x$ (lower panels) in the slice model. $|\chi^{}_z|$ and $V$ are nondimensionalized by $\alpha L$ and $U_\mathrm{QG}$, respectively.  The downward propagating NIWs induce a mean flow change, which slows down the original mean flow.}  
\label{xzfield}
\end{figure}

Figure \ref{xzenergy} shows the evolution of the change in mean energy, wave potential energy and total energy from their initial values in a 14-days simulation. Here, the mean and wave potential energies are the two terms
\begin{equation}
\frac{1}{2} \int \left( \psi_x^2 + \frac{f_0^2}{N^2}\psi_z^2  \right) \, \d \bs{x} \quad \textrm{and} \quad
\frac{N^2}{4} \int \left| \chi^{}_x \right|^2  \, \d \bs{x}
\end{equation}
which make up the constant total energy. The figure confirms that, overall, NIWs act as an energy sink for the mean flow. 
The net energy transfer from mean flow to NIWs is concentrated within the first 5 days; afterwards, the energy exchange is much smaller and its sign alternates. 
The NIW amplitude $|\chi^{}_z|$ and the change in the mean velocity $V=\psi_x$ are shown in Figure \ref{xzfield}. 
Their feedback results in a slowing down of the mean flow, consistent with the energy loss and  collocated with the NIW wavepacket. An important feature of the mean-flow evolution is that it is reversible: at each location, the flow velocity returns to its initial value once the NIWs have propagated away. 
This is a particularity of the slice model, specifically of the diagnostic relation existing between the mean flow and the NIW amplitude. We next consider another two-dimensional model in which the NIW--mean-flow interactions lead to an irreversible behaviour.

\subsection{Vertically plane wave} \label{Sec_tw}

A simple two-dimensional model in the $(x,y)$ plane is obtained by assuming that the wave field takes the form of a plane wave in the vertical, that is, $\chi^{}_z = \varphi(x,y,t) \ex^{\ii m z}$ for some complex function $\varphi$ and vertical wavenumber $m$. This is consistent with a barotropic mean flow $\psi=\psi(x,y,t)$. Introducing this restricted form of the solution into the coupled model (\ref{model}) reduces it to
\begin{subequations}\label{2D1_eq}
\begin{align}
\partial_t \varphi + \partial (\psi,\varphi ) + \ii \beta y \varphi - \frac{\ii N^2}{2m^2f_0} \nabla^2 \varphi +  \frac{\ii}{2}\nabla^2\psi \,\varphi &= 0,\\
\partial_t q + \partial (\psi,q ) &= 0,
\end{align}
\end{subequations}
where 
\begin{eqnarray}
q = \beta y + \nabla^2 \psi + \frac{\ii f_0}{2}\partial(\varphi^*,\varphi) + \frac{f_0}{4} \nabla^2 | \varphi|^2 .
\end{eqnarray}

As an illustration, we consider the propagation of a vorticity dipole in a NIW field on the $f$-plane ($\beta=0$). We carry out simulations initialising the streamfunction $\psi$ to match the vorticity 
\begin{equation} \label{lamb}
\omega = \nabla^2 \psi  =  \left\{
\begin{matrix}
{\displaystyle \frac{2kU}{J_0(\kappa a)} }J_1(\kappa r)\sin\theta, & r<a \\
0, & r>a
\end{matrix} 
\right.,
\end{equation}
of the \citet{Lamb1932} dipole propagating at speed $U$ in the $y$-direction. Here $(r,\theta)$ are polar coordinates,  $a$ characterises the spatial scale of the dipole, $J_n$ are the Bessel functions of the first kind of order $n$, and $\kappa$ is determined by solving the matching condition $J_1(\kappa a)= 0$. 

We carry out a numerical simulation in a periodic domain of size $500 \,\mr{km} \times 500 \,\mr{km}$ using a pseudospectral method. Because of the periodisation, the vorticity (\ref{lamb}) does not exactly correspond to that of a dipole steadily propagating a speed $U$; however, for the dipole size $a=40 \, \mr{km}$ that we take, the differences are minor. We take the other parameters to be $U = 0.05 \, \mr{m s^{-1}}$, $f_0 = 10^{-4} \, \mr{s}^{-1}$, and  $N=0.01 \,\mr{s^{-1}}$. Taking $L=a$ gives a Rossby number $\Ro = 0.0125$.
The initial wave amplitude is chosen as the Gaussian 
\begin{equation}
\varphi = A\ex^{-(k_0 (y-y_0))^2},
\end{equation}
where $A = 1.5 \, \mr{km}$, $k_0 = 2\times 10^{-5} \, \mr{m^{-1}}$ and $y_0=250$ km. This implies that  $\alpha = A/L = 0.0375$ and $U_\mathrm{NIW} = 0.15 \, \mr{m \, s^{-1}}$.
The vertical scale of wave is taken as $m = 0.02 \, \mr{m^{-1}}$, so $\epsilon=0.125$. We therefore have that $\Ro < \alpha < \epsilon\approx\Ro^{1/2}$, consistent with our scaling.
The initial position of the dipole ($r=0$) and wavepacket (maximum of $|\varphi|$) are $(0.5, \,0.3)$ and $y = 0.5$ when  distances are normalised by the domain size of $500\, \mr{km}$.

\begin{figure}
\centering
\includegraphics[width=7.5cm]{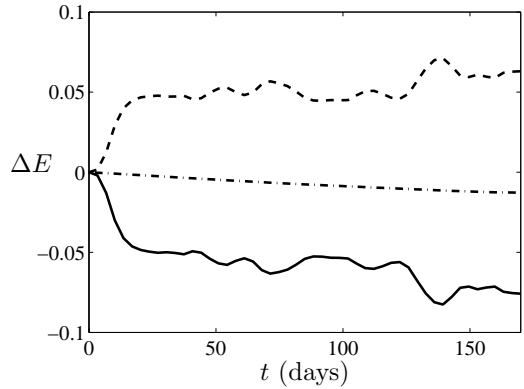}
\caption{Same as Figure \ref{xzenergy} but for the simulation of a vortex dipole propagating in a field of vertically travelling NIWs. {The energy changes are normalized by the initial mean flow energy.}}
\label{dipoleenergy}
\end{figure}

\begin{figure}
\centering
\includegraphics[width=\textwidth]{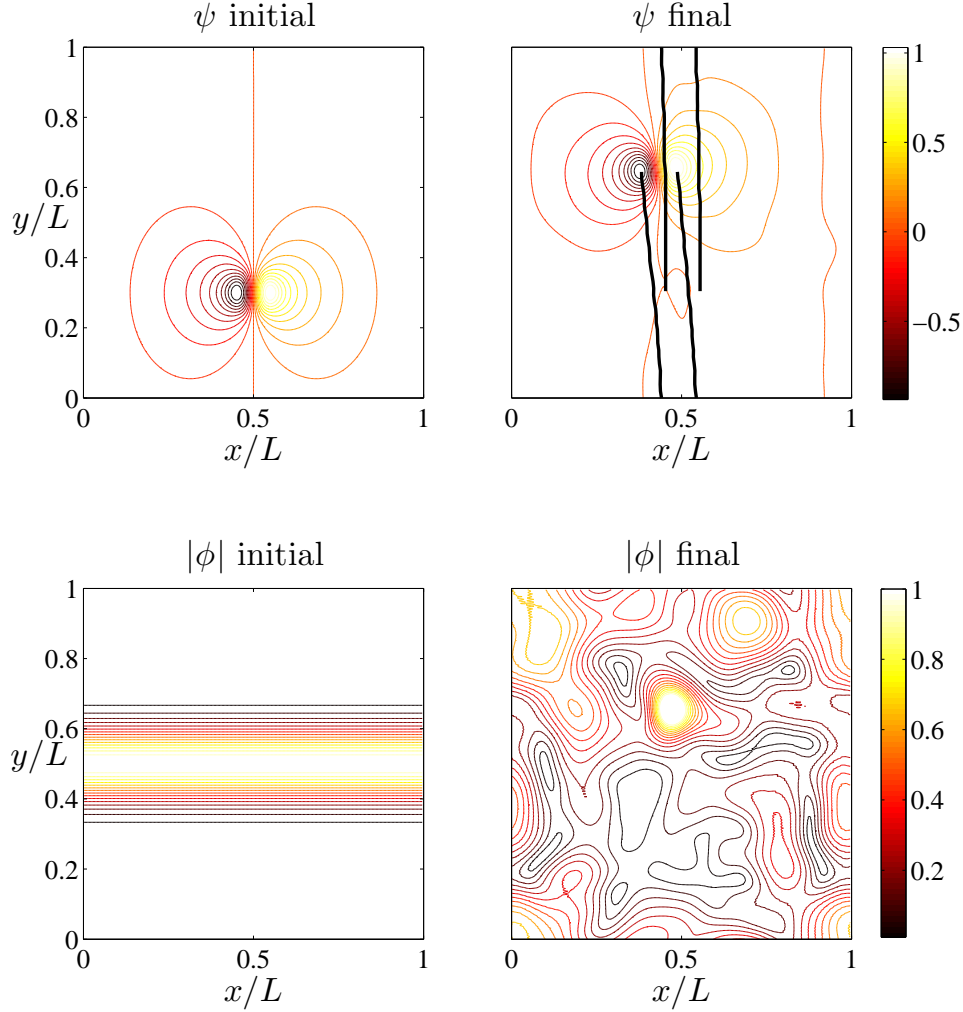}
\caption{NIW-dipole interaction: initial (left panels) and final (right panels) streamfunction $\psi$ (top) and NIW amplitude $|\varphi|$ (bottom). Both $\psi$ and $|\varphi|$ have been normalised by their maximum value at the initial time.
The trajectories of the vorticity maximum and minimum  shown by the thick black lines in the top right panel indicate the motion of the dipole during the simulation (colour online).} 
\label{dipolecontour}
\end{figure}

We report the results of an integration time of $t = 1.5\times 10^7 \, \mr{s} \approx 173 \, \mathrm{days}$, within which the dipole travels about $1\frac{1}{2}$ domain size. The changes in mean and wave energies (normalised by the initial mean energy)  are shown as functions of time in Figure \ref{dipoleenergy}. As in the slice model, the increase of NIW energy is compensated by a loss of mean-flow energy. Using (\ref{aa}) and $\epsilon=0.1$, we can estimate  the relative mean energy change to be about $0.05$, in agreement with the numerical results. 
The initial and final streamfunction $\psi$ and wave amplitude $|\varphi|$ are shown in Figure \ref{dipolecontour}. This also shows the trajectories of the vorticity maximum and minimum as an indication of the dipole's trajectory. The NIWs, {which partly concentrate in the anticyclonic core of the dipole trough a well-established mechanism \citep[e.g.][and references therein]{dani-et-al2}, have an obvious impact
on the mean flow:} instead of propagating in a straight line $x = \mathrm{const}.$, the dipole deforms and is deflected to the left. This illustrates the irreversible nature of the wave--mean flow interactions when, unlike in the slice model, the potential vorticity is not constant.
{The phenomenon is  reminiscent of the deflection of dipoles observed by \citet{snyd-et-al07} in simulations of the spontaneous generation of inertia-gravity waves by dipoles; there is a possible connection that might be worth exploring.}

\section{Discussion} \label{sec:discussion}

In this paper, we derive and study a model of the interactions between slow balanced motion and fast NIWs in the ocean. 
The model is obtained within the GLM framework \citep[e.g.][]{buhl09} or, more precisely, its glm variant \citep{Sowa2010}, and neglects dissipative effects. In its simplest form \eqn{model}, the model consists of the YBJ model of NIW propagation \citep{Youn1997} coupled with a modified quasi-geostrophic equation. As expected from general GLM theory \citep{Buhl1998,buhl09,Salm2013}, the modification consists solely in a change in the relation between streamfunction and potential vorticity which adds to the standard QGPV a quadratic wave contribution.\footnote{A comparison between averaging formalisms (glm, GLM and others) in Appendix \ref{app:altder} shows that this wave contribution arises as the sum  of the curl of a  pseudomomentum, a wave-induced mean-stratification change and a mean-density change, with the exact form of each term depending on the formalism.} Thus NIWs impact the dynamics of  potential vorticity by changing its advection in what is, in general, an irreversible manner. 
The assumption that the waves are near inertial leads to drastic simplifications, reducing the wave part of the dynamics to the YBJ equation for a single (complex) amplitude $\chi$ evolving on the same  time scale as the balanced flow. 

Our YBJ-QG coupled model can be thought of as providing a parameterisation of NIW effects, with the fast NIWs regarded as a subgrid phenomenon in time. In this view, the YBJ is an asymptotically motivated closure for the NIWs: it provides enough information about the NIWs to compute their impact on the balanced flow. We emphasise that the derivation relies on a scale separation in time only and does not assume that the waves have a small spatial scales, unlike previous applications of GLM \citep{Gjaj1996,Buhl1998}. This is crucial for NIWs since they are forced by atmospheric winds at horizontal scales that are much larger than the oceanic mesoscales. It is also practically convenient since the YBJ and QG equations can be solved numerically on the same grid, so that the coupled model requires only about  three times as much computational effort as the standard QG equation. 

As discussed in \S\,\ref{sec:coupledmodel}, the model is not fully consistent asymptotically. This is because the different aspect ratios it assumes for NIWs and balanced motion, specifically $m/k = \eps^{-1} N/f_0 \gg N/f_0$ and $L/H = O(N/f_0)$, cannot be expected to persist: the feedback of the NIWs implies that their aspect ratio is imprinted onto the balanced flow, leading to an increase in $L/H$ and potentially to a breakdown of the assumption of order-one Burger number that underlies the quasi-geostrophic approximation. In practice this may not be significant: the NIWs contribute to the quasi-geostrophic velocity $\nabla^\bot \psi$ through a term that is twice smoother in the vertical than the NIWs amplitude $\chi^{}_z$ itself (because of the Helmholtz inversion in (\ref{mqgpv})). As a result, short vertical fluctuations in $\chi^{}_z$ have a limited impact on $\nabla^\bot \psi$. Furthermore, in the case of locally planar NIWs, it is the envelope scale that is imprinted onto $\nabla^\bot \psi$ rather than the (much shorter) wavelength. 
Finally, the existence of a coupled YBJ--primitive-equation model with  conservations of potential vorticity, energy and action analogous to those of the YBJ-QG model suggests that conclusions inferred from the latter model are robust. {Nonetheless, it might be desirable to treat the difference in the vertical scales $H$ and $m^{-1}$ in a fully consistent way by applying a multiscale method in space as well as in time. It is unclear, however, whether a model derived in this manner would be significantly different from the YBJ-QG model.}

In this paper, we discuss some qualitative aspects of the interactions between balanced flow and NIWs in the ocean, mostly based on the remarkably simple action and energy conservation laws of the YBJ-QG model. The conservation of action  implies the complete absence of spontaneous NIW generation in the model, consistent with the expected exponentially smallness of this phenomenon \citep{Vann2013}. The conservation laws further indicate that NIWs forced at large scales by atmospheric winds provide an energy sink for the oceanic balanced motion through a mechanism that can be termed `stimulated wave generation'. This is potentially significant: several mechanisms have been proposed to explain the dissipation of mesoscale energy but it is far from clear whether they are efficient enough to balance the flux imposed by the energy source (mainly baroclinic instability). We offer a rough estimate of the power extracted from the mean flow by the mechanism we have identified; this suggests that further consideration is worthwhile. More reliable estimates would require intensive numerical simulations of the YBJ-QG or of the primitive equations and are well beyond the scope of this paper.

\medskip

\noindent
\textbf{Acknowledgements.} The authors thank O. B\"uhler, E. Danioux, D. N. Straub, S. M. Taylor, G. L. Wagner and W. R. Young for valuable discussions. W. R. Young proposed the term `stimulated generation' for the mechanism described in \S\,\ref{sec:implications}. This research is funded by the UK Natural Environment Research Council (grant NE/J022012/1). J.-H. X. acknowledges financial support from the Centre for Numerical Algorithms and Intelligent Software (NAIS).

\appendix

\section{glm average} \label{Aglm}

In glm, the map from mean to perturbed positions is written in terms of a divergence-free vector field, $\bs{\nu}(\bX,t)$ say, as
\begin{equation}
\bs{X} + \bs{\xi}(\bX,t) = \ex^{\bs{\nu}} \bX.  \label{glmmap}
\end{equation}
Here the exponential denotes the flow map generated by $\bs{\nu}$; that is, {defining $\bs{x}(s)$  as the solution of
\begin{equation}
\frac{\d}{\d s} \bs{x}(s) = \bs{\nu}\left(\bs{x}(s),t \right), \quad \textrm{where} \ \ \bs{x}(0)=\bs{X}
\end{equation}
and $t$ is regarded as a fixed parameter, $\ex^{\bs{\nu}} \bX = \bs{x}(1)$.} The glm average is then defined by the condition
\begin{eqnarray}
\av{\bs{\nu}} = 0, \label{glm_Av}
\end{eqnarray}
which replaces GLM's condition $\av{\bs{\xi}}=0$ {(\citealt{Sowa2010}; note that we use the symbol $\bs{\nu}$ for the vector field denoted by $\bs{\eta}$ in their paper}). 
The divergence-free property of $\bs{\nu}$ ensures that \eqn{glmmap} preserves volume. For small perturbations $\alpha \ll 1$, it is easy to relate $\bs{\xi}$ to $\bs{\nu}$ order-by-order in $\alpha$. Expanding $\bs{\xi}$ according to \eqn{ExpPer} and, similarly, $\bs{\nu}$ according to $\bs{\nu} =  \bs{\nu}^{(1)} +  \bs{\nu}^{(2)} +\cdots$, we can use \eqn{glmmap} to write
\begin{equation}
\bs{\xi} =  \bs{\nu} +  \frac{1}{2}\bs{\nu} \cdot \nabla_3 \bs{\nu} + \cdots 
=   \bs{\nu}^{(1)} +  \br{ \frac{1}{2}\bs{\nu}^{(1)}\cdot\nabla_3 \bs{\nu}^{(1)} + \bs{\nu}^{(2)} } + \cdots.
\end{equation}
Identifying the first two orders in $\alpha$ yields
\begin{equation}
\bs{\nu}^{(1)} = \bs{\xi}^{(1)} \quad \textrm{and} \quad 
\bs{\nu}^{(2)} = \bs{\xi}^{(2)} - \frac{1}{2}\bs{\nu}^{(1)}\cdot\nabla_3 \bs{\nu}^{(1)}.
\end{equation}
The condition (\ref{glm_Av}) then becomes
\begin{equation}
\av{\bs{\xi}^{(2)}} =  \frac{1}{2} \av{\bs{\nu}^{(1)}\cdot\nabla_3 \bs{\nu}^{(1)}} =  \frac{1}{2}\av{\bs{\xi}^{(1)}\cdot\nabla_3 \bs{\xi}^{(1)}}. \label{glm_2nd}
\end{equation}

\section{Mean dynamics} \label{Mean_Mom}

Following \citet{Salm2013}, the equations governing the mean dynamics are derived from the energy-momentum equations
\begin{eqnarray}\label{EnMoTensor}
\frac{\partial}{\partial {X^j}}\br{ a^i_R \frac{\partial \av{L}}{\partial a^i_{X^j}} } = \frac{\partial \av{L}}{\partial R} - \left.\frac{\partial \av{L}}{\partial R}\right|_{\mr{expl}}^{\chi}
\label{Mean_Cons_Law}
\end{eqnarray}
applied to the density $\av{L}$ associated with the Lagrangian \eqn{MoLa} (i.e.\ $\av{L}$ is the integrand in the expression of $\av{\mathcal{L}}$).
In the energy-momentum equations,  $(X^0,~X^1,~X^2,~X^3) = (T,~X,~Y,~Z)$, $(a^1,a^2,a^3)=(a,b,\theta)$ and Einstein's summation convention is used; $R$ can be taken to be $T$, leading to an energy equation, or $X$, $Y$ or $Z$, leading to the corresponding momentum equations. The sub- and superscript `expl' and $\chi$ attached to the last term in \eqn{EnMoTensor} indicate derivatives of the terms that depend explicitly on $R$, treating the dependence introduced by $\chi$ as such an explicit dependence; in other words, the right-hand side of \eqn{EnMoTensor} collects derivatives associated with the mean flow only.

To keep compact expressions, we make the following definitions:
\begin{subequations} \label{abce}
\begin{align}
A \equiv& \frac{1}{J} \frac{\delta \av{\mc{L}}}{\delta U} =  U  - \br{ f_0Y+\frac{1}{2} \beta Y^2 } + A' \nonumber \\
=& U  - \br{ f_0Y+\frac{1}{2} \beta Y^2 } - \frac{\ii f_0}{4}(\chi_Z \chi^*_{ZX}-\chi^*_Z \chi_{ZX}^{} ) - f_0Y\av{\xi^{(2)}}_X - f_0\av{\eta^{(2)}}, \\
B \equiv& \frac{1}{J} \frac{\delta \av{\mc{L}}}{\delta V} 
= V + B' \nonumber \\
=& V - \frac{\ii f_0}{4}(\chi_Z^{} \chi^*_{ZY}-\chi^*_Z \chi^{}_{ZY} ) - f_0Y\av{\xi^{(2)}}_Y,  \\
C \equiv& \frac{1}{J} \frac{\delta \av{\mc{L}}}{\delta W} = C' \nonumber \\
=& - \frac{\ii f_0}{4}(\chi^{}_Z \chi^*_{ZZ}- \chi^*_Z \chi^{}_{ZZ} ) - f_0Y\av{\xi^{(2)}}_Z,  \\
E \equiv& \frac{\delta \av{L}}{\delta J} 
=\frac{1}{2}(U^2 + V^2 ) - \br{ f_0Y+\frac{1}{2} \beta Y^2 } U  + \theta Z + P + E' \nonumber \\
=&\frac{1}{2}(U^2 + V^2 ) - \br{ f_0Y+\frac{1}{2} \beta Y^2 } U  + \theta Z + P  \nonumber \\
&-\frac{\ii f_0}{4}(\chi_Z^{} D_T \chi_Z^*- \chi_Z^*D_T \chi^{}_Z) -\frac{1}{2}f_0\beta Y|\chi^{}_Z|^2 \nonumber \\
& -f_0YD_T\av{\xi^{(2)}} -f_0\av{\eta^{(2)}}U + \theta \av{\zeta^{(2)}},
\end{align}
\end{subequations}
where $A'$, $B'$, $C'$ and $E'$ group the NIW contributions. Note that $(A',B',C')$ is the wave pseudomomentum. The terms in the energy-momentum tensor (\ref{EnMoTensor}) for $R=T$ can then be written as 
\begin{eqnarray}
a^i_R \frac{\partial \av{L}}{\partial a^i_T} &=&  a^i_R \frac{\partial U^j}{\partial a^i_T} \frac{\partial \av{L}}{\partial U^j} \nonumber \\
&=& - \frac{1}{J} \frac{\partial \av{L}}{\partial U} \jaco{a}{b}{\theta}{R}{Y}{Z} - \frac{1}{J} \frac{\partial \av{L}}{\partial V} \jaco{a}{b}{\theta}{X}{R}{Z} 
-\frac{1}{J} \frac{\partial \av{L}}{\partial W} \jaco{a}{b}{\theta}{X}{Y}{R}  \nonumber \\
&=& - A \jaco{a}{b}{\theta}{R}{Y}{Z} - B \jaco{a}{b}{\theta}{X}{R}{Z} -C \jaco{a}{b}{\theta}{X}{Y}{R} \label{emo1}
\end{eqnarray}
when (\ref{velocity}) is used.
Similarly, for $R=X,\, Y,\,Z$,  we obtain
\begin{subequations} \label{emo2}
\begin{align}
a^i_R \frac{\partial \av{L}}{\partial a^i_X} =&  
- B \jaco{a}{b}{\theta}{R}{T}{Z} -C \jaco{a}{b}{\theta}{R}{Y}{T} + (E-UA-VB-WC)  \jaco{a}{b}{\theta}{R}{Y}{Z},\\
a^i_R \frac{\partial \av{L}}{\partial a^i_Y} =&  
- A \jaco{a}{b}{\theta}{T}{R}{Z} -C \jaco{a}{b}{\theta}{X}{R}{T} + (E-UA-VB-WC)  \jaco{a}{b}{\theta}{X}{R}{Z},\\
a^i_R \frac{\partial \av{L}}{\partial a^i_Z} =&  
- A \jaco{a}{b}{\theta}{T}{Y}{R} -B \jaco{a}{b}{\theta}{X}{T}{R} + (E-UA-VB-WC)  \jaco{a}{b}{\theta}{X}{Y}{R}.
\end{align}
\end{subequations}
Using \eqn{emo1}--\eqn{emo2}, the momentum equations are derived from \eqn{EnMoTensor} with $R=X,\, Y,\, Z$ in the form
\begin{subequations}\label{prieq}
\begin{align}
-D_T A + E_X &= A U_X + B V_X + C W_X + (Z+\av{\zeta^{(2)}}) \theta_X,\\
-D_T B + E_Y &= A U_Y + B V_Y + C W_Y + (Z+\av{\zeta^{(2)}}) \theta_Y,\\
-D_T C + E_Z &= A U_Z + B V_Z + C W_Z + (Z+\av{\zeta^{(2)}}) \theta_Z.
\end{align}
\end{subequations}
Introducing the explicit forms \eqn{abce} of $A,\, B,\, C$ and $D$ leads, after simplifications, to \eqn{meanpri}.

\section{Alternative derivation} \label{app:altder}

In this Appendix we show that the QGPV equation (\ref{MQGPV_Eq}) can be obtained directly from potential-vorticity conservation. In this procedure GLM, glm and indeed, any definition of the average $\av {\bxi^{(2)}}$ gives the same leading order dynamics because the associated mean flow maps are $O(\alpha^2)$ close. The wave contributions to the mean dynamics come from different sources depending on the definition of the average, but their total effect is the same.

We start from the  general Lagrangian (\ref{LA}). Taking $\delta P$ variation we obtain
\begin{equation}
J = 1 + \nabla_3 \cdot \br{\av{\bs{\xi}^{(2)}} - \frac{1}{2}\av{\bs{\xi}^{(1)}\cdot\nabla \bs{\xi}^{(1)}}}.\label{Jacobian}
\end{equation}
The relabeling symmetry of Lagrangian (\ref{LA}) gives potential-vorticity conservation
\begin{equation} \label{df}
D_T\br{ \frac{\nabla \theta \cdot \nabla \times \bs{A}}{J} } = 0,
\end{equation}
where $\bs{A} = (A,\,B,\,C)$ are defined as in (\ref{abce}) but with the Lagrangian (\ref{LA}) in place of \eqn{MoLa} \citep{Salm2013}.

Under quasi-geostrophic scaling and using the buoyancy equation (\ref{geo3}) to replace $W$ in the above equation, we obtain
\begin{equation}
D_T^0 \br{ \frac{N^2(B_X-A_Y) + f_0\theta'_Z}{J}} - \frac{f_0}{N^2}D_T^0(\theta'(N^2)_Z) = 0,
\end{equation}
where $\theta$ follows the definition  (\ref{theta}).
By substituting
\begin{equation}
B_X-A_Y = f_0 +\beta Y + \nabla^2 \psi  + \frac{\ii f_0}{2}\partial(\chi_Z^*,\chi^{}_Z) + f_0 \av{\partial_x \xi^{(2)} + \partial_y \eta^{(2)}}, \label{aas}
\end{equation}
and (\ref{Jacobian}), we obtain the modified QGPV equation 
\begin{equation}
D_T^0\br{f_0 +\beta Y + \nabla^2 \psi + \partial_Z\left(\frac{f_0^2}{N^2}\partial_Z\psi\right) + \frac{\ii f_0}{2}\partial(\chi_Z^*,\chi^{}_Z) +  \frac{f_0}{2}{\nabla\cdot}\av{\bs{\xi}^{(1)}\cdot\nabla_3\bs{\xi}^{(1)}}}=0, \label{QGPV_eq}
\end{equation}
identical to (\ref{MQGPV_Eq}) since the last term is equal to $f G(\chi^*,\chi)$. Note that the cancellation of the second-order mean displacements (term $\nabla_3\cdot\av{\bs{\xi}^{(2)}}$) indicates that this equation is independent of the specific averaging used to define the Lagrangian mean. In contrast, the individual wave contributions to the QGPV, namely the curl of the pseudomomentum (wave terms in (\ref{aas})), the buoyancy term $N^2 f_0 \av{\partial_Z \zeta^{(2)}}$ and the density correction (divergence in (\ref{Jacobian})) depend on the averaging used. 
{A relation reducing to (\ref{QGPV_eq}) for NIWs was derived by \citeauthor{Holm2011} (2011; their Eq.\ (3.10)) using GLM theory.}

\bibliographystyle{jfm}
\bibliography{niwref}

\begin{thebibliography}{45}
\expandafter\ifx\csname natexlab\endcsname\relax\def\natexlab#1{#1}\fi

\bibitem[Andrews \& McIntyre(1978)]{Andr1978}
{\sc Andrews, D.~G. \& McIntyre, M.~E.} 1978 An exact theory of nonlinear waves
  on a {L}agrangian-mean flow. {\em J. Fluid Mech.\/} {\bf 89}~(4), 609--646.

\bibitem[Badulin \& Shrira(1993)]{Badu1993}
{\sc Badulin, S.~I. \& Shrira, V.~I.} 1993 On the irreversibility of
  internal-wave dynamics due to wave trapping by mean flow inhomogeneities.
  {P}art 1. {L}ocal analysis. {\em J. Fluid Mech.\/} {\bf 251}, 21--53.

\bibitem[Balmforth {\em et~al.\/}(1998)Balmforth, Llewellyn-Smith \&
  Young]{Balm-et-al1998}
{\sc Balmforth, N.J., Llewellyn-Smith, S.~G. \& Young, W.~R.} 1998 Enhanced
  dispersion of near-inertial waves in an idealised geostrophic flow. {\em J.
  Mar. Res.\/} {\bf 56}, 1--40.

\bibitem[Berestetskii {\em et~al.\/}(1982)Berestetskii, Lifshitz \&
  Pitaevskii]{quantum}
{\sc Berestetskii, V.~B., Lifshitz, E.~M. \& Pitaevskii, L.~P.} 1982 {\em
  Quantum electrodynamics\/}, 2nd edn. Cambridge University Press.

\bibitem[Bokhove {\em et~al.\/}(1998)Bokhove, Vanneste \& Warn]{Bokh1998}
{\sc Bokhove, O., Vanneste, J. \& Warn, T.} 1998 A variational formulation for
  barotropic quasi-geostrophic flows. {\em Geophys. Astrophys. Fluid Dynam.\/}
  {\bf 88}, 67--79.

\bibitem[B\"uhler(2009)]{buhl09}
{\sc B\"uhler, O.} 2009 {\em Waves and mean flows\/}. Cambridge University
  Press.

\bibitem[B\"uhler \& McIntyre(1998)]{Buhl1998}
{\sc B\"uhler, O. \& McIntyre, M.~E.} 1998 On non-dissipative wave--€"mean
  interactions in the atmosphere or oceans. {\em J. Fluid Mech.\/} {\bf 354},
  301--343.

\bibitem[B\"uhler \& McIntyre(2005)]{Buhl2005}
{\sc B\"uhler, O. \& McIntyre, M.~E.} 2005 Wave capture and wave-vortex
  duality. {\em J. Fluid Mech.\/} {\bf 534}, 67--95.

\bibitem[Cotter \& Reich(2004)]{Cotter-Reich}
{\sc Cotter, C.~J. \& Reich, S.} 2004 Adiabatic invariance and applications:
  From molecular dynamics to numerical weather prediction. {\em BIT Numer.\
  Math.\/} {\bf 44}, 439--455.

\bibitem[Danioux {\em et~al.\/}(2015)Danioux, Vanneste \&
  B{\"u}hler]{dani-et-al2}
{\sc Danioux, E., Vanneste, J. \& B{\"u}hler, O.} 2015 On the concentration of
  near-inertial waves in anticyclones. {\em J. Fluid Mech.\/} Submitted.

\bibitem[Danioux {\em et~al.\/}(2012)Danioux, Vanneste, Klein \&
  Sasaki]{dani-et-al}
{\sc Danioux, E., Vanneste, J., Klein, P. \& Sasaki, H.} 2012 Spontaneous
  inertia-gravity-wave generation by surface-intensified turbulence. {\em J.
  Fluid Mech.\/} {\bf 699}, 153--157.

\bibitem[D'Asaro {\em et~al.\/}(1995)D'Asaro, Eriksen, Levine, Paulson, Niiler
  \& Meurs]{DAsa1995}
{\sc D'Asaro, E.~A., Eriksen, C.~C., Levine, M.~D., Paulson, C.~A., Niiler, P.
  \& Meurs, P.~Van} 1995 Upper-ocean inertial currents forced by a strong
  storm. {P}art {I}: Data and comparisons with linear theory. {\em J. Phys.
  Oceanogr.\/} {\bf 25}, 2909--2936.

\bibitem[Duhaut \& Straub(2006)]{Duha2006}
{\sc Duhaut, T. H.~A. \& Straub, D.~N.} 2006 Wind stress dependence on ocean
  surface velocity: Implications for mechanical energy input to ocean
  circulation. {\em J. Phys. Oceanogr.\/} {\bf 36}, 202--211.

\bibitem[Falkovich {\em et~al.\/}(1994)Falkovich, Kuznetsov \&
  Medvedev]{Falk1994}
{\sc Falkovich, G., Kuznetsov, E. \& Medvedev, S.} 1994 Nonlinear interaction
  between long inertio-gravity and {R}ossby waves. {\em Nonlin. Processes
  Geophys.\/} {\bf 1}, 168--171.

\bibitem[Ferrari \& Wunsch(2009)]{Ferr2009}
{\sc Ferrari, R. \& Wunsch, C.} 2009 Ocean circulation kinetic energy:
  reservoirs, sources, and sinks. {\em Annu. Rev. Fluid Mech.\/} {\bf 31},
  962--971.

\bibitem[Fu(1981)]{Fu1981}
{\sc Fu, L.-L.} 1981 Observations and models of inertial waves in the deep
  ocean. {\em Rev. Geophys. Space Phys.\/} {\bf 19}, 141--170.

\bibitem[Garrett(2001)]{Garr2001}
{\sc Garrett, C.} 2001 What is the {``}near-inertial{"} band and why is it
  different from the rest of the internal wave spectrum? {\em J. Phys.
  Oceanogr.\/} {\bf 41}, 253--282.

\bibitem[Gertz \& Straub(2009)]{Gert2009}
{\sc Gertz, A. \& Straub, D.~N.} 2009 Near-inertial oscillations and the
  damping of midlatitude gyres: a modeling study. {\em J. Phys. Oceanogr.\/}
  {\bf 39}, 2338--2350.

\bibitem[Gjaja \& Holm(1996)]{Gjaj1996}
{\sc Gjaja, I. \& Holm, D.~D.} 1996 Self-consistent {H}amiltonian dynamics of
  wave mean-flow interaction for a rotating stratified incompressible fluid.
  {\em Physica D\/} {\bf 98}, 343--378.

\bibitem[Goldstein(1980)]{gold80}
{\sc Goldstein, H.} 1980 {\em Classical mechanics\/}, 2nd edn. Addison-Wesley.

\bibitem[Grimshaw(1984)]{Grim1984}
{\sc Grimshaw, R.} 1984 Wave action and wave-mean flow interaction, with
  application to stratified shear flows. {\em Ann. Rev. Fluid Mech.\/} {\bf
  16}, 11--44.

\bibitem[Holliday \& McIntyre(1981)]{holl-mcin}
{\sc Holliday, D. \& McIntyre, M.~E.} 1981 On potential energy density in an
  incompressible stratified fluid. {\em J. Fluid Mech.\/} {\bf 107}, 221--225.

\bibitem[Holm {\em et~al.\/}(2009)Holm, Schmah \& Stoica]{Holm2009}
{\sc Holm, D.~D., Schmah, T. \& Stoica, C.} 2009 {\em Geometric mechanics and
  symmetry\/}. Oxford University Press.

\bibitem[Holmes-Cerfon€ {\em et~al.\/}(2011)Holmes-Cerfon€, {B{\"u}hler} \&
  Ferrari]{Holm2011}
{\sc Holmes-Cerfon€, M., {B{\"u}hler}, O. \& Ferrari, R.} 2011 Particle
  dispersion by random waves in the rotating boussinesq system. {\em J. Fluid
  Mech.\/} {\bf 670}, 150--175.

\bibitem[Hunter \& Ifrim(2013)]{Hunt2013}
{\sc Hunter, J.~K. \& Ifrim, M.} 2013 A quasi-linear {S}chr\"odinger equation
  for large amplitude inertial oscillations in a rotating shallow fluid. {\em
  IMA J. Appl. Math.\/} {\bf 78}, 777--796.

\bibitem[Kunze(1985)]{Kunz1985}
{\sc Kunze, E.} 1985 Near-inertial wave propagation in geostrophic shear. {\em
  J. Phys. Oceanogr.\/} {\bf 15}, 544--565.

\bibitem[Lamb(1932)]{Lamb1932}
{\sc Lamb, H.} 1932 {\em Hydrodynamics\/}, 6th edn. Cambridge University Press.

\bibitem[Medvedev \& Zeitlin(1997)]{Medv2007}
{\sc Medvedev, S.B. \& Zeitlin, V.} 1997 Turbulence of near-inertial waves in
  the continuously stratified fluid. {\em Physics Letters A\/} {\bf 371}~(3),
  221--227.

\bibitem[Mooers(1975{\natexlab{{\em a\/}}})]{Mooers1975a}
{\sc Mooers, C. N.~K.} 1975{\natexlab{{\em a\/}}} Several effects of a
  baroclinic current on the cross-stream propagation of inertial-internal
  waves. {\em Geophys. Fluid Dyn.\/} {\bf 6}, 245--275.

\bibitem[Mooers(1975{\natexlab{{\em b\/}}})]{Mooers1975b}
{\sc Mooers, C. N.~K.} 1975{\natexlab{{\em b\/}}} Several effects of baroclinic
  currents on the three-dimensional propagation of inertial-internal waves.
  {\em Geophys. Fluid Dyn.\/} {\bf 6}, 277--284.

\bibitem[Nikurashin {\em et~al.\/}(2013)Nikurashin, Vallis \&
  Adcroft]{Niku2013}
{\sc Nikurashin, M., Vallis, G.~K. \& Adcroft, A.} 2013 Routes to energy
  dissipation for geostrophic flows in the {S}outhern {O}cean. {\em Nature
  Geoscience\/} {\bf 6}, 48--51.

\bibitem[Oliver(2006)]{Oliv2006}
{\sc Oliver, M.} 2006 Variational asymptotics for rotating shallow water near
  geostrophy: a transformational approach. {\em J. Fluid Mech.\/} {\bf 551},
  197--234.

\bibitem[Salmon(1988)]{Salm1988}
{\sc Salmon, R.} 1988 Hamiltonian fluid mechanics. {\em Ann. Rev. Fluid
  Mech.\/} {\bf 20}, 225--256.

\bibitem[Salmon(2013)]{Salm2013}
{\sc Salmon, R.} 2013 An alternative view of generalized {L}agrangian mean
  theory. {\em J. Fluid Mech.\/} {\bf 719}, 165--182.

\bibitem[Shepherd(1990)]{shep90}
{\sc Shepherd, T.~G.} 1990 Symmetries, conservation laws and {H}amiltonian
  structure in geophysical fluid dynamics. {\em Adv. Geophys.\/} {\bf 32},
  287--338.

\bibitem[Snyder {\em et~al.\/}(2007)Snyder, Muraki, Plougonven \&
  Zhang]{snyd-et-al07}
{\sc Snyder, C., Muraki, D., Plougonven, R. \& Zhang, F.} 2007 Inertia-gravity
  waves generated within a dipole vortex. {\em J. Atmos. Sci.\/} {\bf 64},
  4417--4431.

\bibitem[Soward \& Roberts(2010)]{Sowa2010}
{\sc Soward, A.~M. \& Roberts, P.~H.} 2010 The hybrid {E}uler--{L}agrange
  procedure using an extension of {M}offatt's method. {\em J. Fluid Mech.\/}
  {\bf 661}, 45--72.

\bibitem[Vallis(2006)]{vall06}
{\sc Vallis, G.~K.} 2006 {\em Atmospheric and oceanic fluid dynamics:
  fundamentals and large-scale circulation\/}. Cambridge University Press.

\bibitem[Vanneste(2013)]{Vann2013}
{\sc Vanneste, J.} 2013 Balance and spontaneous generation in geophysical
  flows. {\em Annu. Rev. Fluid Mech.\/} {\bf 45}, 147--172.

\bibitem[Vanneste(2014)]{Vann2014}
{\sc Vanneste, J.} 2014 Deriving the {Y}oung-{B}en {J}elloul model of
  near-inertial waves by {W}hitham averaging. {\em arXiv:1410.0253\/} .

\bibitem[Whitham(1974)]{Whit1974}
{\sc Whitham, G.~B.} 1974 {\em Linear and nonlinear waves\/}. Wiley, 636 pages.

\bibitem[Wunsch \& Ferrari(2004)]{Wuns2004}
{\sc Wunsch, C. \& Ferrari, R.} 2004 Vertical mixing, energy, and the eneral
  circulation of the oceans. {\em Annu. Rev. Fluid Mech.\/} {\bf 36}, 281--314.

\bibitem[Young \& {Ben Jelloul}(1997)]{Youn1997}
{\sc Young, W.~R. \& {Ben Jelloul}, M.} 1997 Propagation of near-inertial
  oscillations through a geostrophic flow. {\em J. Mar. Res.\/} {\bf 55}~(4),
  735--766.

\bibitem[Young {\em et~al.\/}(2008)Young, Tsang \& Balmforth]{Youn2008}
{\sc Young, W.~R., Tsang, Y.-K. \& Balmforth, N.~J.} 2008 Near-inertial
  parametric subharmonic instability. {\em J. Fluid Mech.\/} {\bf 607}, 25--49.

\bibitem[Zeitlin {\em et~al.\/}(2003)Zeitlin, Reznik \& {Ben
  Jelloul}]{Zeitlin2003}
{\sc Zeitlin, V., Reznik, G.~M. \& {Ben Jelloul}, M.} 2003 Nonlinear theory of
  geostrophic adjustment. {P}art 2. {T}wo-layer and continuously stratified
  primitive equations. {\em J. Fluid Mech.\/} {\bf 491}, 207--228.

\end{thebibliography}

\end{document}